\documentclass[12pt,preprint]{aastex}
\shorttitle{SiO Polarization Variability}
\shortauthors{Glenn et al.}
\begin{document}
  \newcommand{\ignore}[1]{}
  \newcommand{\kms}[0]{km $ \textup{s}^{-1} $}
  \title{
    A POLARIZATION SURVEY OF SiO MASER VARIABILITY IN EVOLVED STARS
  }
  \author{J. Glenn\altaffilmark{1}, P.R. Jewell\altaffilmark{2},
          R. Fourre\altaffilmark{1}, and L. Miaja\altaffilmark{1}
          }
  \altaffiltext{1}{CASA, University of Colorado, 389-UCB
                   Boulder, CO 80309, USA}

  \altaffiltext{2}{National Radio Astronomy Observatory, P.O. Box 2,
                   Green Bank, WV 24944.  The National Radio Astronomy
		Observatory is a facility of the National Science 
		Foundation operated uncer cooperative agreement by 
		Associated Universities, Inc.}
  \begin{abstract}

We have monitored the SiO ($v = 1$, ${\mbox J} = 2 \rightarrow 1$)
maser polarization in 17 variable stars (Miras, OH-IR stars, and
supergiants) to investigate the long-term persistence of masers.  The
8 epochs of observations span 2.5 years, thereby sampling multiple
cycles for these stars with typical periods of $\sim 1$ year.  The
average polarization was 23\% with a typical dispersion of 7\%,
although the variability differed substantially from star to star.  In
the Stokes $q$ and $u$ spectra of individual stars, a few strong maser
features tended to dominate the polarization, with the maser features
persisting for less than one stellar cycle for some stars, and for
multiple cycles for a few stars.  Because individual masers are not
resolved in our beam-averaged total intensity spectra, we correlated
the polarization spectra between epochs to measure the characteristic
lifetimes of the features, rather than attempting to trace the
evolution of separate line components.  We found that individual maser
feature lifetimes ranged from a few months or less to $> 2$ years.
These data indicate that for the sample of stars as a whole, the
masers are not reset at a particular stellar phase.

  \end{abstract}
  \section{Introduction}
    \label{sec:intro}

SiO masers are found in a variety of evolved giant and super-giant
stars, including Miras, semi-regular variables, M super-giants, and
OH/IR stars.  Giant and supergiant stellar atmospheres and their mass
loss mechanisms are not fully understood: convection and composition
variations are complex, and detailed atmospheric models are just
emerging.  The location of SiO masers, close to the photospheres,
makes them useful probes of the extended atmospheres and the mass loss 
phenomenon.

Spectroscopy of evolved stars has shown significant variability in
maser intensities and line widths
\citep{cla82,cla84,bar85,jew91,her98}.  $^{28}$SiO (hereafter simply
SiO) masers have been observed in rotational transitions from
\mbox{$J=1\rightarrow0$} to \mbox{$J=7\rightarrow6$} and from
vibrational states \mbox{$v=0$} to \mbox{$v=3$}, from frequencies of
43 GHz to 302 GHz \citep{jew87,pij94,cho96,gra97}. VLBI observations
have revealed that the masers are generally located in ring structures
in the circumstellar envelopes
\citep{dia94,miy94,gre95,hum96,col96,bob97,doe98}.  Ring radii
typically range from 1.5 to 4 stellar radii.  Since the observed rings
present themselves perpendicular to the line of sight, it has been
postulated that the masing takes place on the surfaces of spheres
centered on the stars, and that the masing radiation escapes
tangentially to the spheres.  Combined with the tangential
polarization vectors, this structure has been taken as evidence for
radiative pumping of the masers (Desmurs et al. 2000).

Previous observations of some Miras indicated that individual maser
spectral features appeared to persist throughout entire stellar cycles
(e.g., Clark et al. 1984). However, the continuity of maser features
was also found to be disrupted at various times throughout the stellar
cycles, including at maximum light (e.g., Clark, Troland, \& Miller
1985) and at SiO maser brightness minimum (Mart\a'inez, Bujarrabal, \&
Alcolea 1998). It has been proposed that the masers are created anew
by periodic shock waves propagating through the extended atmospheres
and lagging the optical maximum by about $\sim0.2 $ periods
\citep{cla84}.  Recent numerical models have been used to
simulate the SiO maser variability with a constant infrared radiation
field combined with shocks (Humphreys et al. 2002).

To investigate the relationship between maser variability and stellar
cycle, we observed the 86.2434 GHz SiO (\mbox{$v=1$},
\mbox{$J=2\rightarrow1$}; antenna temperature and polarization)
transition in 17 evolved stars during eight epochs over two and a half
years.  Our observations are not densely sampled enough to investigate
the detailed evolution of masers in any star within a stellar cycle.
However, the large survey size, with sampling up to four stellar
cycles per star, enables us to quantify the long-term polarization
behavior of the masers, complementary to detailed studies of
individual stars.  \S
\ref{sec:obs} describes our observations and \S \ref{sec:data}
presents the results and discussion.  Our conclusions are summarized
in \S \ref{sec:conc}.

  \section{Observations}
    \label{sec:obs}
 
The observations were made with the facility polarimeter and 3-mm
receivers at the NRAO 12 Meter Telescope on Kitt Peak in standard
position-switching mode.  A parallel-wire grid and mirror combination,
with a tuned separation acting as a half-waveplate, modulated the
polarization.  The waveplate was stepped to 16 positions per rotation,
allowing for four independent measurements of polarization per
rotation.  Systematic, instrumental polarization was measured with
observations of planets, and found to be small ($<<1$\%) compared to
the observed stellar polarizations.  The orthogonal senses of
polarization were split by another parallel-wire grid and transmitted
to two receivers.  The polarization efficiency and position angle were
calibrated with a linearly-polarized noise source mounted at the
center of the subreflector.  The parallactic angle correction was
verified with observations of the Crab Nebula.  We used two
128-channel filter bank backends for each receiver, with 100 kHz and
250 kHz resolutions, corresponding to 348 and 870 m s$^{-1}$,
respectively.  The 250 kHz spectra were used only to cross-check the
100 kHz spectra.  Each star was observed one or more times per day,
weather permitting, during the following eight epochs: 1995 April,
1995 October, 1996 March, 1996 July, 1996 September, 1997 February,
1997 June, and 1997 October. Calibration was done on the $T_{R}^\ast$
scale.  The conversion factor to flux density is 32
Jy/K$(T_{R}^\ast)$. The stellar phases and periods were derived from
Kholopov (2002) and the AAVSO data base
(http://www.aavso.org/adata/onlinedata/).  Throughout this paper, the
stellar phases are referred to the optical maxima.  The spectra and
polarizations for each star for each epoch are displayed in Figures 8 to 24..

  \section{Data Analysis and Discussion}
    \label{sec:data}

  \subsection{Periodicity of SiO Brightness}
    \label{subsec:int}

The peak line temperatures vary dramatically, up to nearly a factor of
100 for R Hya from stellar phase 0.15 to 0.97.  Previous observations
have revealed periodic maser line temperature variations matched to
the stellar periods but with a phase lag of $ \sim0.2 $ periods
\citep{cla84, gra97}.  To test for this signal in our data, for each
star we summed the antenna temperatures in each backend channel over
the entire spectral line, normalized the sums to the maximum
observered line temperature, and plotted them versus the stellar phase
(Figure 1).  For the ensemble, the integrated line temperatures are
clearly periodic.  A least-squares best-fit sine wave fits the
aggregate-normalized maser temperatures well with a phase lag of
$\Delta \phi = 0.11$ and an uncertainty $\sigma_{\Delta \phi} = 0.03$,
in broad agreement with the previous results, but tending toward a
slightly smaller phase shift.

  \subsection{General Characteristics of Maser Polarizations}
    \label{subsec:pol}

The Stokes parameters were calculated for each velocity channel for
each observation (Figures 8 through 25): $ I $, $ q $ (N-S polarized
${\mbox T_R^\ast}$ minus E-W polarized ${\mbox T_R^\ast}$), and $ u $
(NE-SW polarized ${\mbox T_R^\ast}$ minus NW-SE polarized ${\mbox
T_R^\ast}$), with polarization $ P $ \mbox{$ = {\sqrt{q^{2}+u^{2}}}/I
$} and position angle $\theta = {1\over2}tan^{-1}({U\over Q})$, where
$Q = q/I$ and $U = u/I$ are the normalized Stokes parameters and $I =
{\mbox T_R^\ast}$.  Since we are interested in the long term
variability, we averaged the Stokes spectra for each star for the
multiple observations at each epoch (separated by no more than a
couple of days).

The spectra are complex with several line components visible in most
of them.  In general, the polarizations are dominated by a few
highly-polarized components that can be distinguished by their $q$'s
and $u$'s.  Since our observations beam-average all the masers within
each stellar envelope, it is possible that multiple masers contribute
to each observed feature at a given velocity.  Overlapping features
could explain the smooth variations in polarization and position angle
(given by changing relative contributions from $q$ and $u$) across the
spectra.  For example, R Aqr is dominated by 1-3 components that tend
to blend together for all the epochs in our survey.  In contrast, $o$
Ceti, VY CMa, U Her, R Leo, and VX Sgr have separate line components
that can be distinguished even in the absence of polarization
information.  Under the assumption that the masing cells are
distributed in a ring-like manner with tangential polarization
vectors, the line blending of multiple components will reduce the
observed polarization.  Thus, our interpretation of independent masers
leads to a lower limit to the actual polarization of independent masing
entities.

As a broad characterization of the maser polarization, we averaged the
polarization across the spectral lines for each star for each epoch.
The percentage polarizations are shown as a function of Julian Day in
Figure 2.  The spectrum-averaged polarization varies substantially for
most of the stars.  However, stars with high average polarization
tended to retain a high polarization: $o$ Ceti ($\langle P \rangle =
44$\%, rms $= 9$\%) and S CrB ($\langle P \rangle = 41$\%, rms $=
8$\%).  Stars with moderate polarization tended to retain moderate
polarization, and stars with low polarization tended to retain low
polarization, e.g., VX Sgr ($\langle P \rangle = 5$\%, rms $= 2$\%;
see Figure 3, where $\sigma_P/\langle P \rangle <1$ for all of the
stars).  The average polarization for all the stars over all the
epochs was 23\% (Table 1).

To look for evidence of cyclic polarization variation, the
polarization is plotted in Figure 4 versus the stellar phase $\phi$.
No significant systematic polarization variability is apparent, with
the possible exception of a minimum at $\phi = 0.2$, approximately
corresponding to the peak in maser brightness.  This minimum at $\phi
= 0.2$ is suspect: a single observation of 40\% polarization at $\phi
= 0.2$ would make the apparent minimum invisible.  The most highly
polarized star, $o$ Ceti was not observed near $\phi = 0.2$, and if it
had been the visual impression of a minimum likely would be absent.

A subtle shift in polarization as a function of stellar phase can be
tested by binning the polarizations for the ensemble into the high
intensity phase (corresponding to the brightest half of each cycle for
SiO; $0.85 < \phi < 0.35$ in Figures 1 and 4) and the low intensity
phase ($0.35 <
\phi < 0.85$).  Excluding VY CMa, which is aperiodic, and
OH2.6-0.4, for which we did not find a single stellar maximum, the
average high intensity phase polarization was 23\% ($\sigma = 4$\%;
standard deviation of the mean) and the average low intensity phase
polarization was 23\% ($\sigma = 3$\%; standard deviation of the
mean).  Thus there is no convincing evidence of systematic
polarization variability in the high intensity phase compared to the
low intensity phase.

The integrated line temperatures were higher during the high intensity
phase than during the low intensity phase by an average factor of 1.3.
Under the assumption that the masers are distributed on circumstellar
rings with tangential polarization vectors, we can relate the
integrated line temperatures and percentage polarizations.  An
increased number of masers with a constant average brightness (leading
to the increased high intensity phase brightness) would tend to
decrease the beam-averaged polarizations by a factor of $\sqrt{n}$ (in
the limit of large $n$), where $n$ is the number of independent
masers.  For a brightness increase of 1.3, the polarization would be
expected to fall by a factor of 1.14 from 23\%, during the low
intensity phase, to 20\% during the high intensity phase (for
$n\propto{\mbox T_R^\ast}$). Since the percentage polarization is
actually is equal during both phases at 23\% and the $1\sigma$
uncertainty in the difference is 5\%, the interpretation is
inconclusive.  The spectra themselves do not show a clear dependence
on the number of line polarization features as a function of stellar
phase, although polarization spectral features should not be assigned
to discrete masers due to the beam averaging.

The spectrum-averaged polarization position angles $\theta$ do not
have random distributions for many of the stars in our sample (Figure
5).  This either indicates: 1) that there is some long-term
($\Delta\phi > 1$) persistence in the masers, or 2) if the masers are
distributed in circumstellar rings with tangential polarization
vectors, that particular regions in the circumstellar envelopes tend
to consistently mase most brightly for $\Delta\phi > 1$, or both.  R
Aqr, VY CMa, $o$ Ceti, and S CrB, for example, have position angles
that slowly vary around mean values, with position angle dispersions
$\sigma_\theta$ of $30^\circ$, $5^\circ$, $20^\circ$, and $19^\circ$,
respectively.  VY CMa is dominated by just a few small features that
remain coherent for the entire span of our observations.  So, for
these stars, even if the discrete masers do not remain coherent for
greater than one period, the masing is dominated by one or two
(opposing) quadrants of the stellar disks.  However, this is not true
for all the stars: the mean $\sigma_\theta = 49^\circ$, indicating
that dominant maser coherence is not maintained for $\Delta\phi > 1$
in general.

  \subsection{Estimation of Maser Lifetimes}
    \label{subsec:dec}

To quantitatively test the prediction that masers are reborn each
stellar cycle with the passage of a shock wave, we compared the
polarization spectral feature lifetimes, which are assumed to be
representative of the maser lifetimes, with the stellar periods.  To
avoid the ambiguities associated with the decomposition of a
continuous spectrum into a discrete sum of spectral components, we
used a statistical method to estimate the feature lifetimes.  In
addition to numerical simplicity, this approach takes in account all
maser polarization, even from features too small to be individually
extracted. Although all maser emission in the spectra contribute
to the derived lifetimes, brighter components contribute more than
fainter ones.  This technique does not require assigning individual
masers to spectral polarization features on a one-to-one basis.

For computational convenience, each spectrum was treated as a complex
function $S(\nu)$ of velocity, where $q(\nu)$ was the real part and
$u(\nu)$ was the imaginary part (similarly, the polarization could be
represented in polar coordinates: $P = P_0 e^{i\theta}$).  We write
the correlation $ C_{ij} $ between the spectra of two epochs $i$ and
$j$, observed at times $T_i$ and $T_j$, as
    \begin{equation}
      C_{ij} = \frac
               {
                 \displaystyle
                 \int_{\nu_{L}}^{\nu_{U}}{
                   S_{i}(\nu) \cdot S_{j}(\nu) \, d\nu
                 }
               }
               {\left[
                 \displaystyle
                 \int_{\nu_{L}}^{\nu_{U}}{|S_{i}(\nu)|^{2} d\nu}
                 \int_{\nu_{L}}^{\nu_{U}}{|S_{j}(\nu)|^{2} d\nu}
               \right]^{1/2}}.
    \end{equation}
The time between the two observations $i$ and $j$ is $\Delta T_{ij} =
T_j - T_i$, and the integrals are numerical sums over the spectral
lines binned in velocity by the receiver backends.  $C_{ij}$ would be
unity for $i = j$, and one expects the $C_{ij}$'s to become smaller as
$\Delta T_{ij}$ increases and the masers evolve, come, and go.  To
estimate an average maser lifetime for a star, the width of $(\Delta
T_{ij}$, $C_{ij})$ distribution for all $ij$ pairs was fit to a gaussian expectation
function $H(\Delta T)$ with width (in time) given by $\sigma$.  With
this formalism, the width in the correlation function is
$\sqrt{2}\sigma$, and we take the average spectral feature lifetime
to be $2\sigma$.  The average spectral component lifetime is not the
lifetime of a maser, but represents a characteristic time over which the $q$ and $u$
spectral features persist.

The stars were sampled from one to six times per period (typically 2-3
times per period), with an average of 6.4 observations per star over
30 months, and a typical time between observations of 3.75 months.
Therefore, lifetimes shorter than a few months or greater than 30
months cannot be measured.  The uncertainties in the fits of the
correlation distribution to the expectation function were on the order
of 1-2 weeks, but the systematic errors in applying this technique to
poorly sampled data should dominate the uncertainties.

Average maser lifetimes derived with this technique range from 85 to
800 days, with the lowest number corresponding well to the minimum
time between observations.  The lifetimes are plotted versus the
stellar periods in Figure 6; the stellar periods cluster at just over
one year.  Long lifetimes were derived for $o$ Ceti (545 days) and S
CrB (500 days).  Inspection of their spectra in Figures 12 and 14,
respectively, reveals several strong polarization features that
persist for many of the observations, even for $>1$ stellar period.  R
Leo, with a derived lifetime of 415 days, also exhibits the
persistence of some features, enabling a robust lifetime to be
derived.  VY CMa is aperiodic, and hence does not appear in Figure 6,
but does have weak polarization features at $\sim 0$ km/s that
persisted from 1996 March to 1997 October, and one feature at $\sim
-1.5$ km/s that is present in all the epochs, consistent with the 645
day derived lifetime.  For the majority of the stars in the sample,
the characteristic lifetimes are shorter than the stellar periods,
indicating that many masers do not survive entire stellar cycles.  The
OH-IR stars and supergiants do not seem to differ from the Miras in
their maser lifetimes.

To summarize, one rule does not apply to all of the stars: some
exhibit maser line features that persist for more than one stellar
cycle, while others exhibit masers that do not persist for even a
single full cycle.  Beyond this, our survey does not have sufficient
temporal resolution to determine whether most masers go through a
``reset'' at any particular phase.

  \subsection{Maser Line Widths}
    \label{subsec:velo}

In a manner similar to the maser lifetime estimation, the maser line
widths can be estimated in km/s without resorting to fitting multiple
line components, which are not well constrained for low-level spectral
features.  We derived polarization line feature widths from the
autocorrelation function $ A(\Delta \nu) $ of the polarization
spectrum $S(\nu)$:
\begin{equation}
A(\Delta \nu) = \int_{\nu_{L}}^{\nu_{U}}{S(\nu)S(\nu+\Delta \nu) d\nu}.
\end{equation}
The $A(\Delta \nu)$ closely approximate Gaussians: in the limit that the
spectrum $S(\nu)$ were the sum of Gaussian features of width $\sigma$,
the autocorrelation function would be a Gaussian of width $
\sqrt{2}\sigma $.  We defined the average velocity dispersion as the
dispersion that would produce an autocorrelation function best fitting
the observed result $A(\Delta \nu)$.  

Averaged over the ensemble, the velocity dispersion is 0.52 \kms.  The
range of widths is small, 0.33 to 0.67 \kms, suggesting that a common
physical process leads to an upper limit for the maser
velocity-widths, or that blending occurs on this scale due to beam
averaging.  Figure 7 shows that the maser feature line widths are
independent of stellar phase, indicating that coherent masers are not
velocity broadened by the stellar winds or pulsations during the
stellar cycles.

  \section{Conclusions}
    \label{sec:conc}

We monitored the intensity and polarization of SiO masers in 17
evolved stars, over 2.5 years, to investigate the relationship between
maser variability and stellar phase.  The polarization variability is
not uniform throughout the sample.  Sustained polarizations of nearly
50\% were observed observed in $o$ Ceti, whereas a sustained
polarization of $\sim5$\% was observed in VX Sgr, and the
sample-average polarization was 23\% with a typical dispersion of 7\%.
For the ensemble of stars, the polarization is not related to stellar
phase, with the possible exception of a slight minimum in polarization
at $\phi \sim 0.2$.

In the Stokes $q$ and $u$ spectra of individual stars, a few strong
maser features tended to dominate the polarization.  In some cases,
these masers persisted for several epochs ($>1$ stellar cycle), and in
other cases they evolved much more quickly, with little resemblance
between the spectra from subsequent epochs.  We used a statistical
technique to quantify the lifetimes of maser line polarization
features (which may not correspond to independent masers in our
beam-averaged observations) for comparison to the stellar periods.
For each star, we measured the correlation between the spectra in all
pairs of epochs as a function of the time between the epochs.  This
technique has the advantage of incorporating all of the polarization
information in the spectra, without having to fit individual line
components and trace them from epoch-to-epoch.  For regular variable
stars, we found that the characteristic maser lifetimes ranged from a
few months (the minimum lifetime that could be measured) to 800 days.
These data indicate that the masers are not reset by the passage of
shocks every stellar cycle from stellar pulsations for some of the
stars, and that the maser lifetimes are often shorter than the stellar
periods for others.

We also found a long-term persistence in the spectrum-averaged
polarization position angles for some of the stars.  This is in broad
agreement with the persistence of individual masers, or a preference
for masing in single or opposing quadrants of the stellar disks
(assuming ring-like distributions of masers with tangential
polarizations).  Finally, the widths of the maser line components in
the spectra ranged from $\sim 0.3-0.7$ km/s, with no dependence on
stellar phase.

  \section{Acknowledgments}
    \label{sec:ack}
    We thank the AAVSO International Database and its observers
    worldwide for the dates of stellar maxima.  We thank Phil Maloney for
	helpful discussions and Chris Walker for contributions to the design
	of the polarimeter.

\clearpage

    \begin{deluxetable}{cccccccc}
	\tabletypesize{\scriptsize}     
        \tablewidth{0pt}
        \tablehead{
	\colhead{Star} & \colhead{Stellar Type} & \colhead{Period (Days)} &
	\colhead{$\langle P(\%) \rangle $} &
	\colhead{$\sigma_{\mbox P}(\%)$} & \colhead{$\langle \theta(^\circ) \rangle $} &
	\colhead{$\sigma_\theta(^\circ)$} & \colhead{Characteristic Maser Lifetime (Days)}}
      \startdata
        R Aqr & Mira/symbiotic & 387 & 18.7 &  5.2 & -41 & 30 & 210 \\
        RX Boo    & Mira       & 340 & 34.3 & 16.5 & 51  & 39 & 105\\
        TX Cam    & Mira       & 557 & 14.0 & 2.5  & 0   & 70 & 360 \\
        R Cas     & Mira       & 430 & 23.2 & 8.5  & 15  & 69 & 260 \\
        $o$ Cet   & Mira       & 332 & 44.3 & 9.4  & 0   & 20 & 545 \\
        VY CMa    & Supergiant & ... & 4.5  &  0.5 & 43  & 5  & 645 \\
        S CrB     & Mira       & 360 & 40.7 & 8.1  & 55  & 19 & 500 \\
        $\chi$ Cyg   & Mira    & 408 & 17.4 & 8.3  & 34  & 42 & 160 \\
        RU Her    & Mira       & 485 & 24.0 & 7.0  & 7   & 47 & 85  \\
        U Her     & Mira       & 406 & 17.7 & 3.0  & 7   & 59 & 310 \\
        R Hya     & Mira       & 389 & 32.0 & 17.5 & 9   & 71 & 160 \\
        W Hya     & Mira       & 361 & 16.6 & 7.9  & 35  & 27 & 125 \\
        R Leo     & Mira       & 310 & 22.9 & 4.7  & -19 & 65 & 415 \\
        VX Sgr    & Supergiant & 732 & 5.3  & 2.1  & 11  & 78 & 235 \\
        IK Tau    & OH-IR      & 470 & 12.0 & 4.5  & 22  & 60 & 315 \\
        IRC+10011 & OH-IR      & 660 & 15.3 & 5.3  & 35  & 45 & 160 \\
        OH2.6-0.4 & OH-IR      & ... & 27.2 & 4.3  & -1  & 86 & 800 \\
      \enddata
    \end{deluxetable}

\clearpage

    \begin{figure}
    \plotone{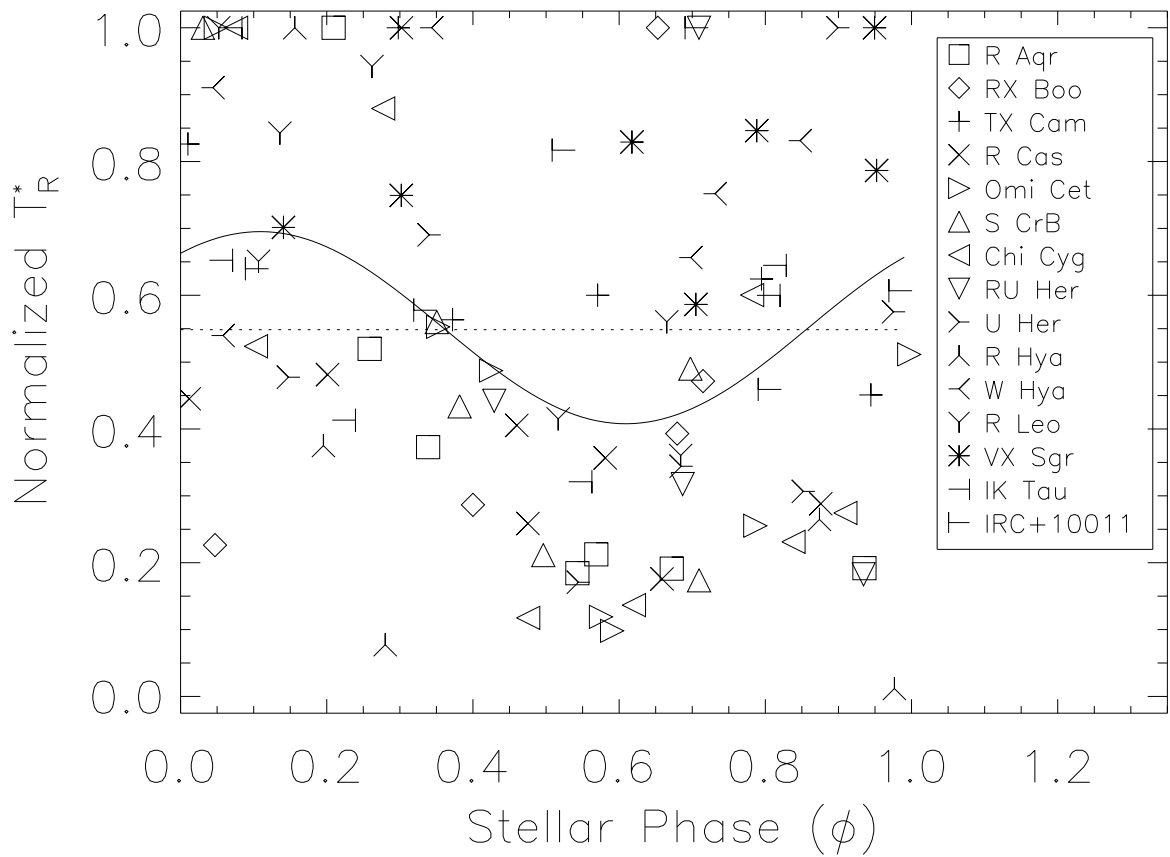}
      \caption{
SiO line temperatures for the survey stars as a function of stellar phase.
To facilitate comparison between the stars, the line temperatures, ${\mbox T_R^*}$, have been normalized for each
star to the maximum observed line temperature for all of the epochs.  The dotted
line shows the mean for the ensemble, and the solid curve is the
best-fit sine wave.  The peak SiO brightness for the average is at a
phase of 0.11, lagging the peak optical brightness.
      }
    \end{figure}
    \placefigure{fig:fig1}

\clearpage

    \begin{figure}
	\includegraphics[scale=0.8]{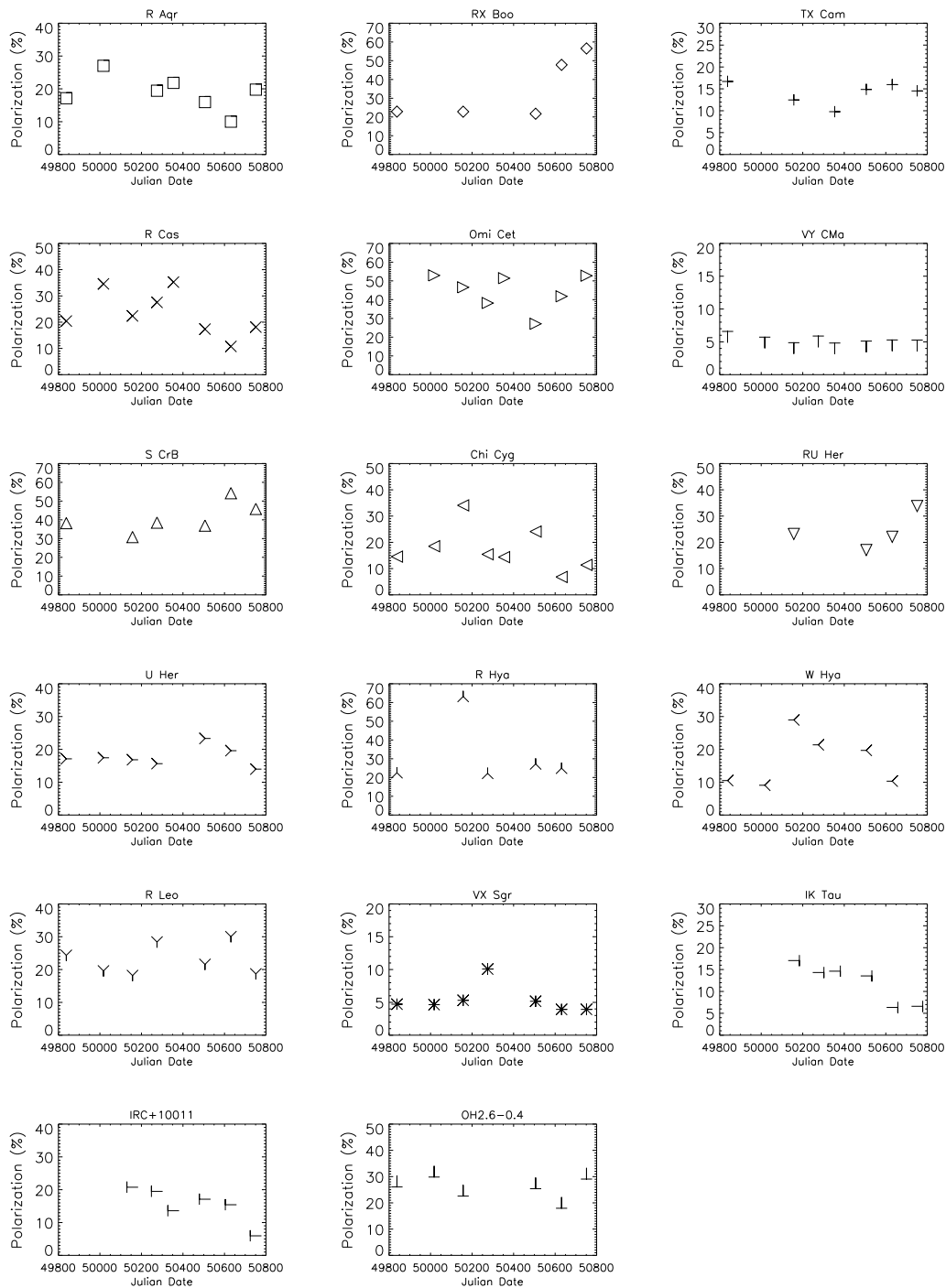}
      \caption{
	Line-integrated percentage polarization as a function of Julian Date. The full
	Julian Date for the first epoch of observations is 2449838.
	The highest polarization for R Hya has very low S/N, which
	leads to an erroneously high polarization.  }
    \end{figure}
    \placefigure{fig:fig2}

 \clearpage

    \begin{figure}
	\plotone{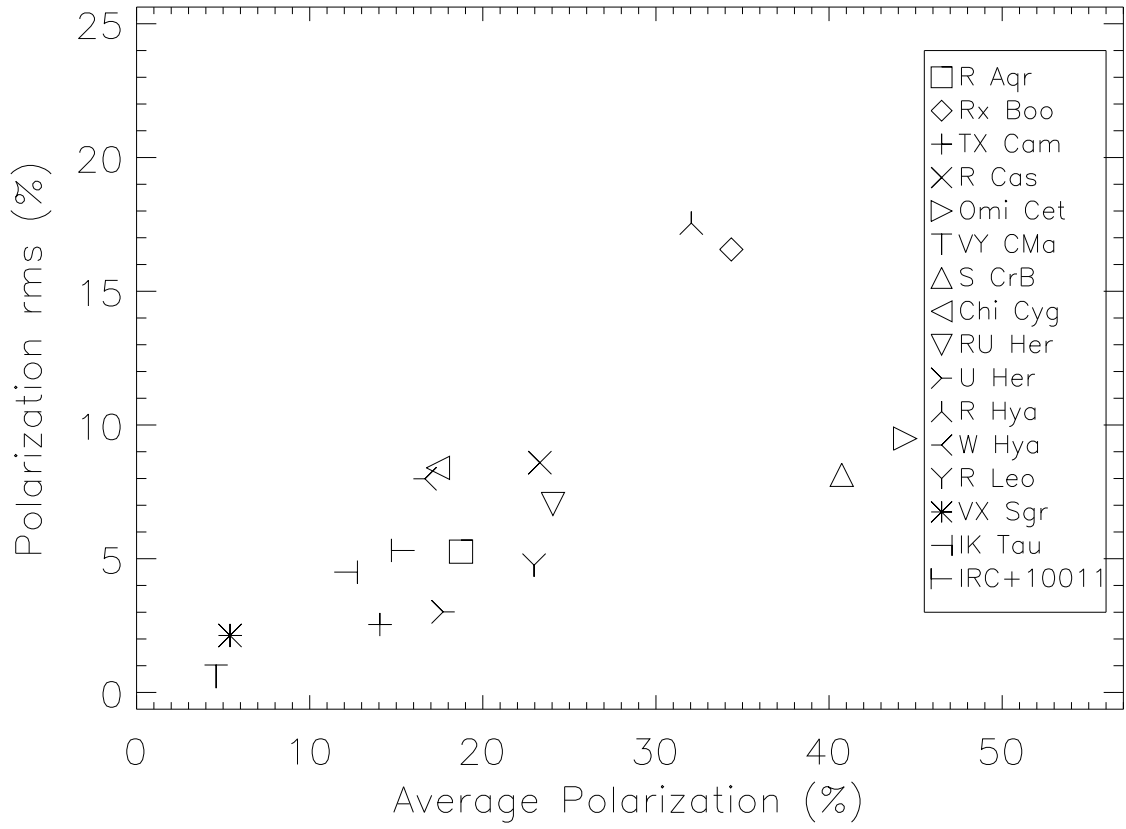}
      \caption{
	Line-integrated epoch-to-epoch polarization rms versus percentage polarization averaged over all the epochs.}
    \end{figure}
    \placefigure{fig:fig3}

\clearpage

   \begin{figure}
	\plotone{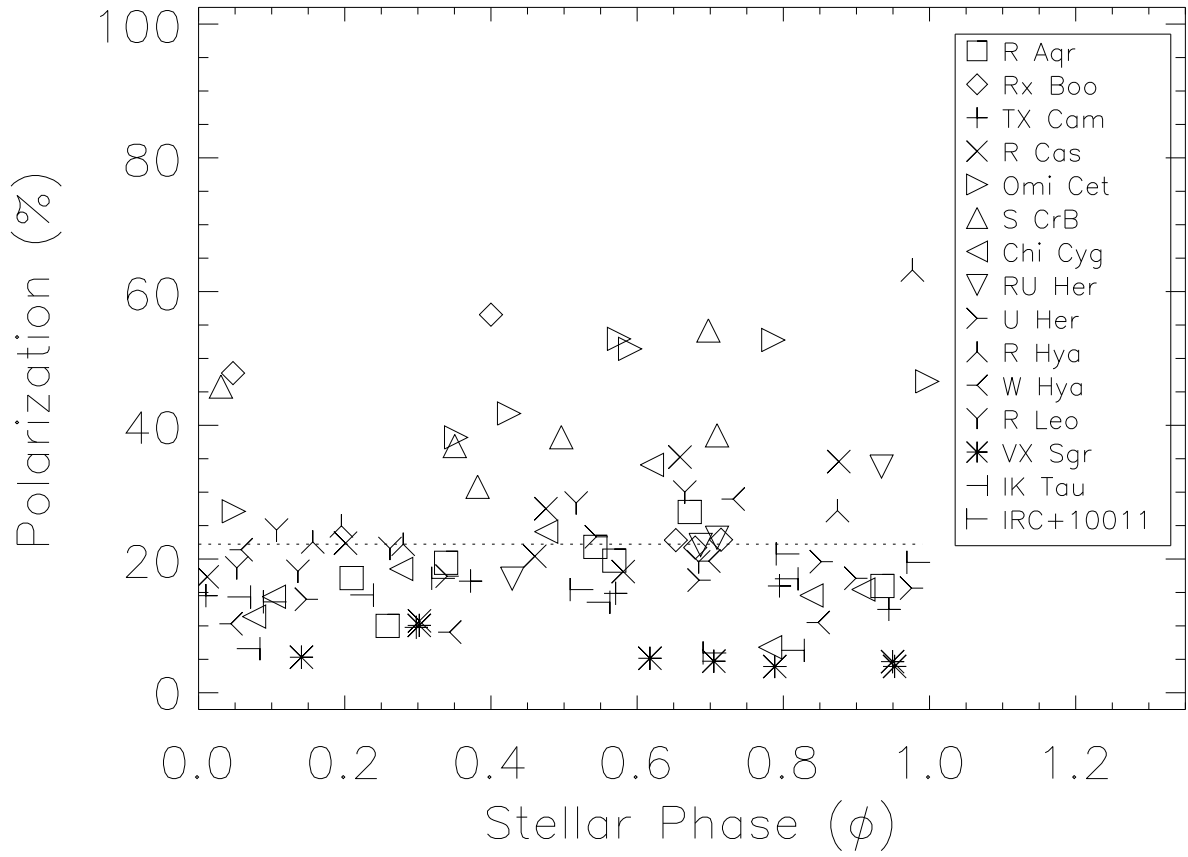}
     \caption{
	Line-integrated polarization as a function of stellar phase.}
    \end{figure}
    \placefigure{fig:fig4}

 \clearpage

    \begin{figure}
     \includegraphics[scale=0.8]{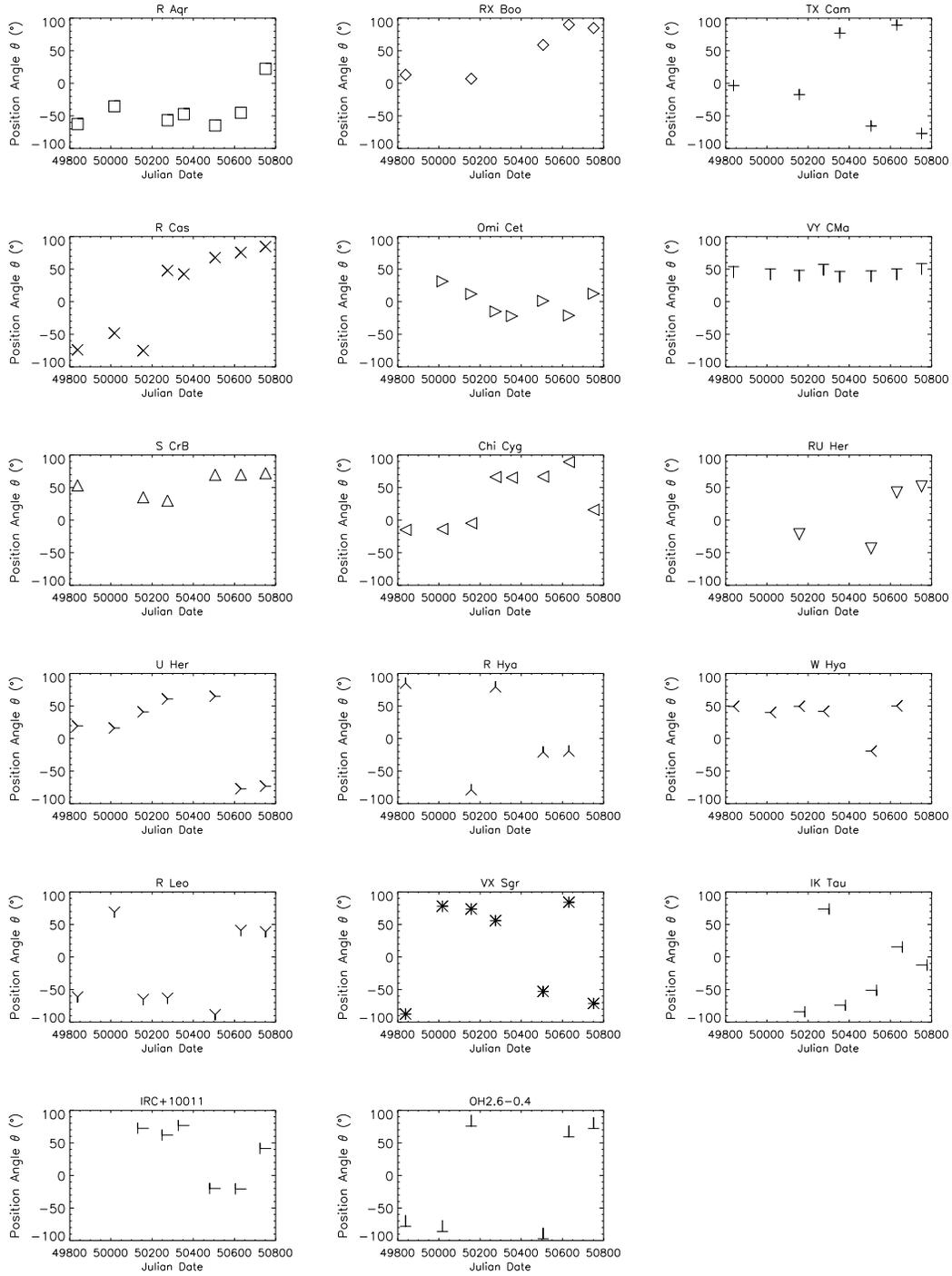}
      \caption{
	Line-integrated polarization position angle as a function of Julian Date. The full
	Julian Date for the first epoch of observations is 2449838.}
    \end{figure}
    \placefigure{fig:fig5}

\clearpage

    \begin{figure}
	\plotone{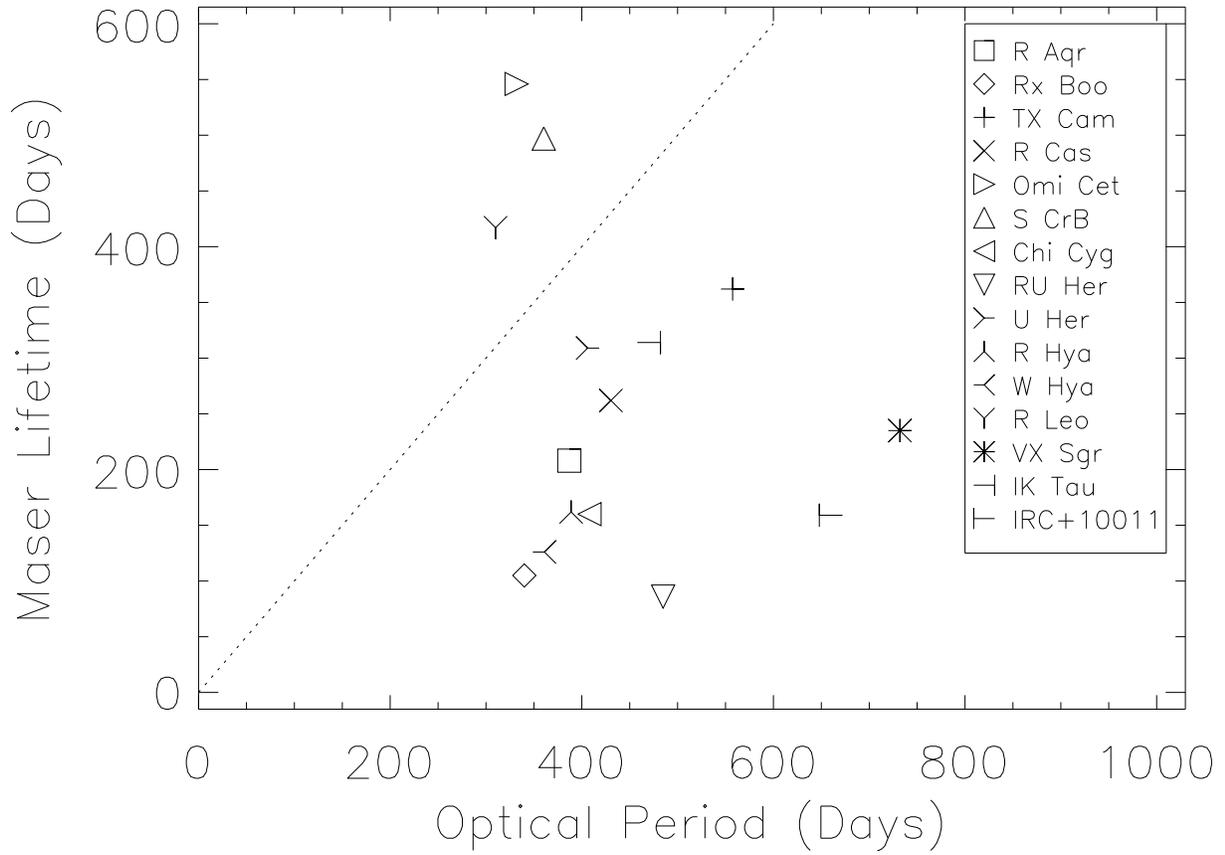}
      \caption{
	Characteristic maser spectral line feature lifetime versus optical period for
	all the periodic stars in the survey.  The maser lifetimes
	were derived by correlating all the pairs of spectra at
	different epochs for each star.  The dotted line has a slope of 
	one; it is meant to show that most of the derived lifetimes are
	shorter than the stellar periods.}
    \end{figure}
    \placefigure{fig:fig6}

    \begin{figure}
      \plotone{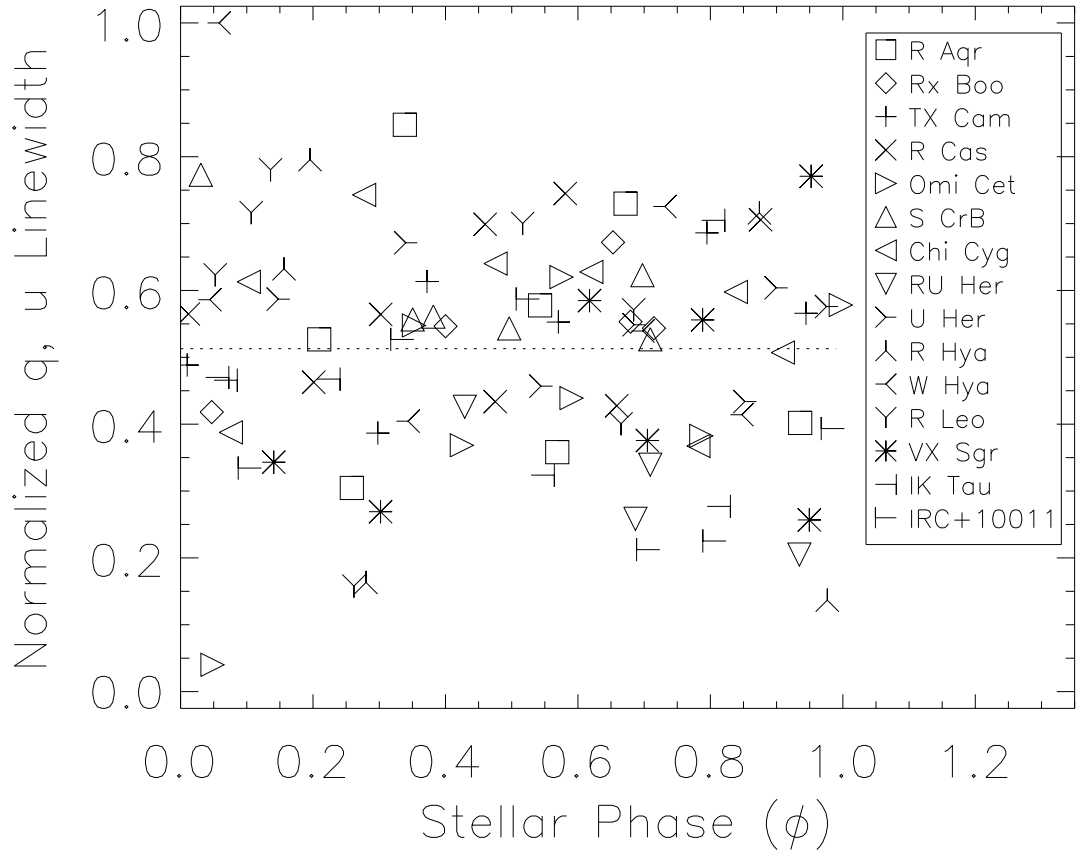}
      \caption{
	Maser polarization feature linewidths versus stellar phase.  
The linewidths were derived from the widths of the autocorrelation function
of the $q$, $u$ spectra for each epoch.  For 
each star, the line width in km/s at each epoch was normalized by the maximum
observed linewidth.}
    \end{figure}
    \placefigure{fig:fig7}

\clearpage

    \begin{figure}
      \includegraphics[scale=0.8]{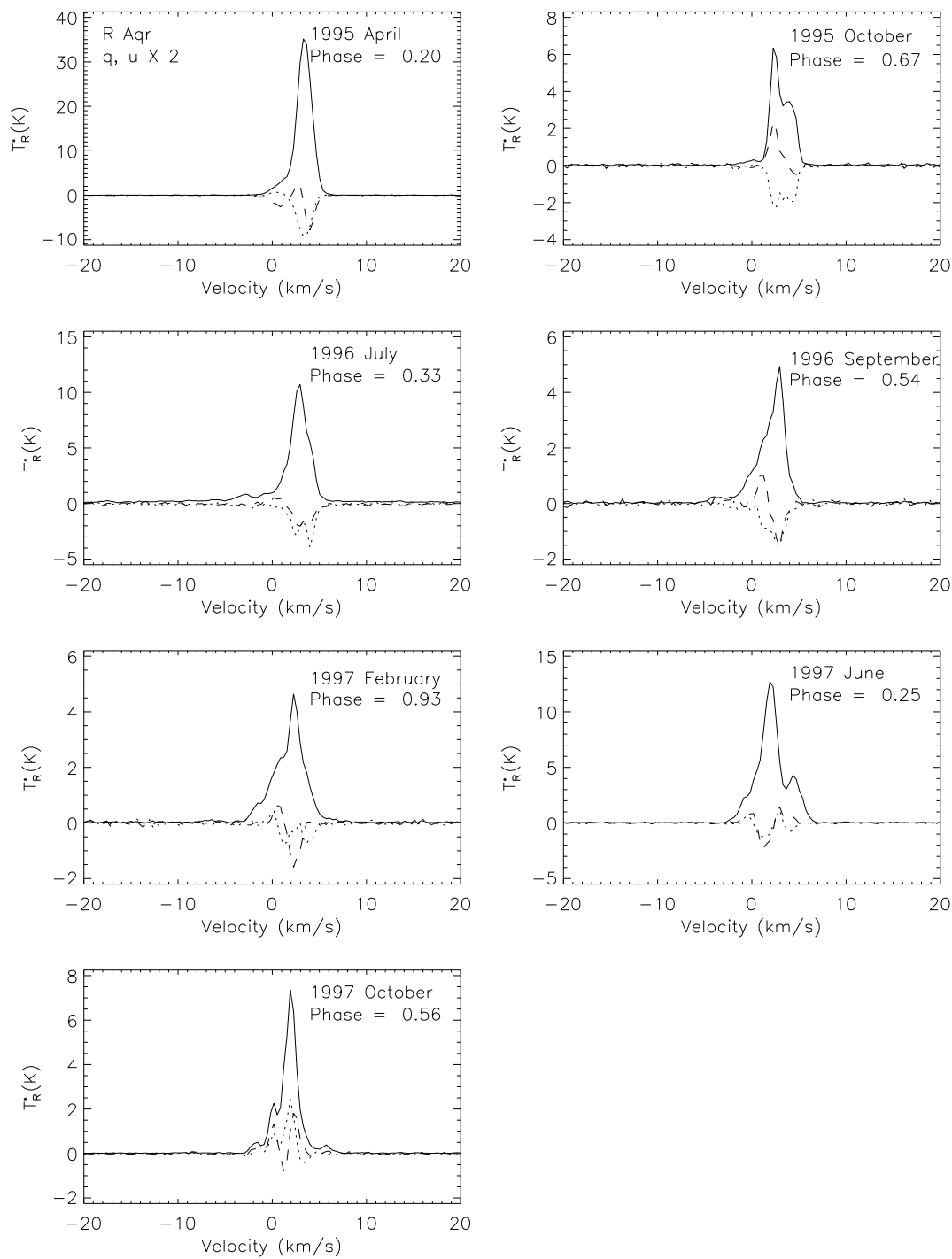}
      \caption{
SiO line temperature and polarization spectra for R Aqr.  There is one
panel per epoch in which the star was observed.  The phase for each
epoch is given.  The non-normalized Stokes parameters, $q$ (dashed
lines) and $u$ (dotted lines), have been multiplied by two in all of
the panels.}
    \end{figure}
    \placefigure{fig:R_Aqr}

\clearpage

    \begin{figure}
      \includegraphics[scale=0.8]{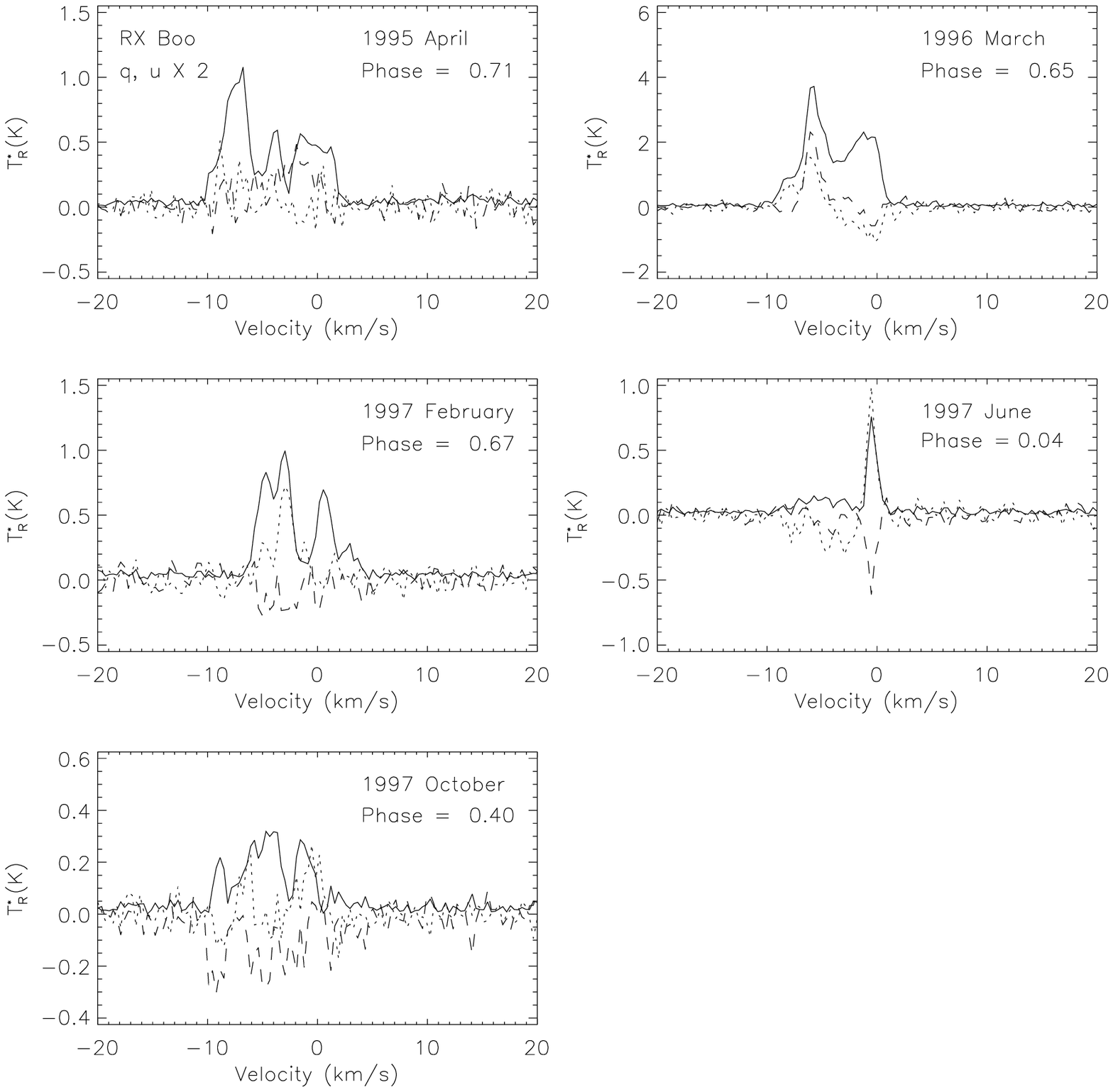}
      \caption{
SiO line temperature and polarization spectra for RX Boo.  The
non-normalized Stokes parameters, $q$ (dashed lines) and $u$ (dotted
lines), have been multiplied by two in all of the panels.}
    \end{figure}
    \placefigure{fig:RX_Boo}

\clearpage

    \begin{figure}
      \includegraphics[scale=0.8]{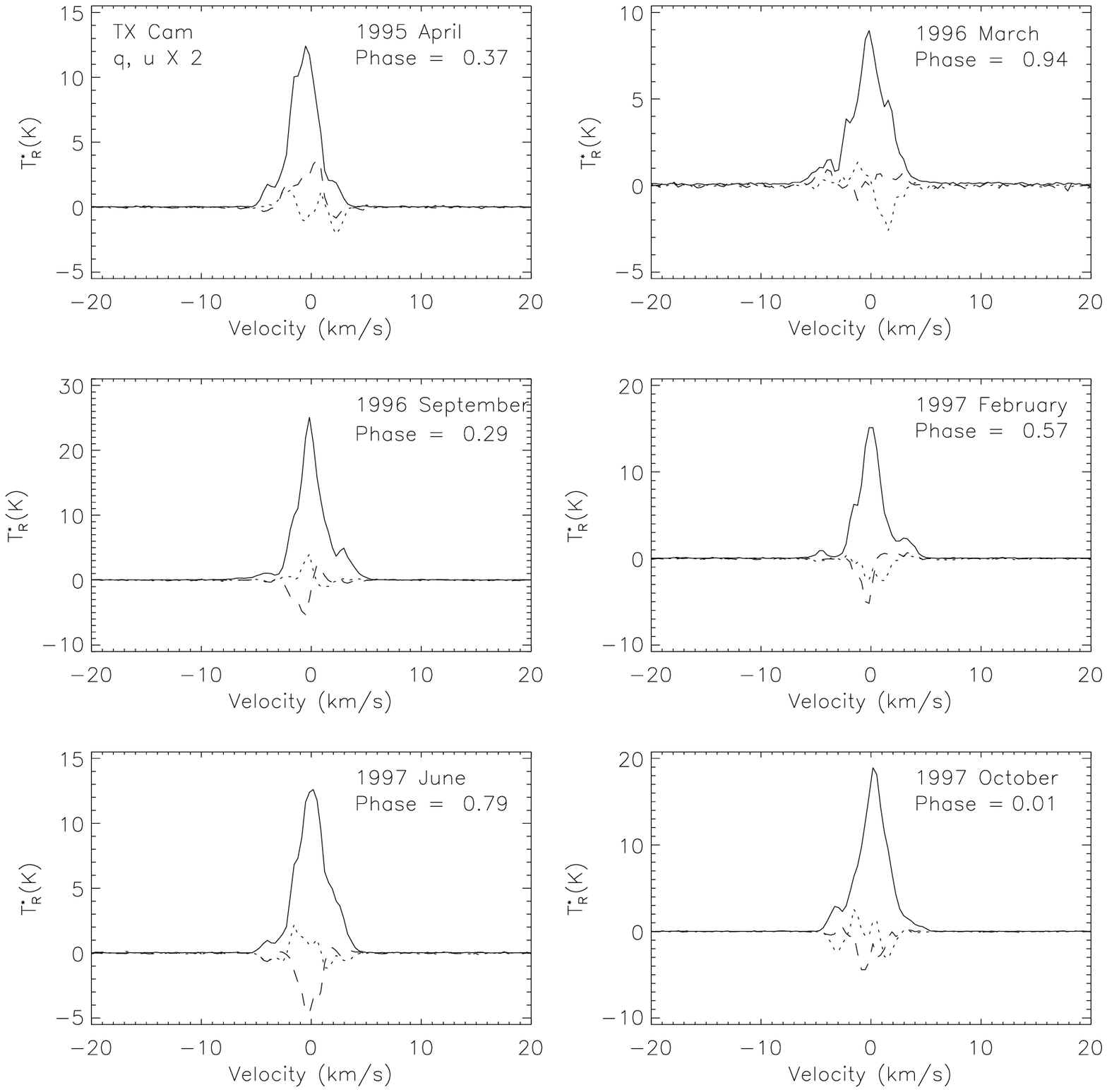}
      \caption{
SiO line temperature and polarization spectra for TX Cam.  The
non-normalized Stokes parameters, $q$ (dashed lines) and $u$ (dotted
lines), have been multiplied by two in all of the panels.}
    \end{figure}
    \placefigure{fig:TX_Cam}

\clearpage

    \begin{figure}
      \includegraphics[scale=0.8]{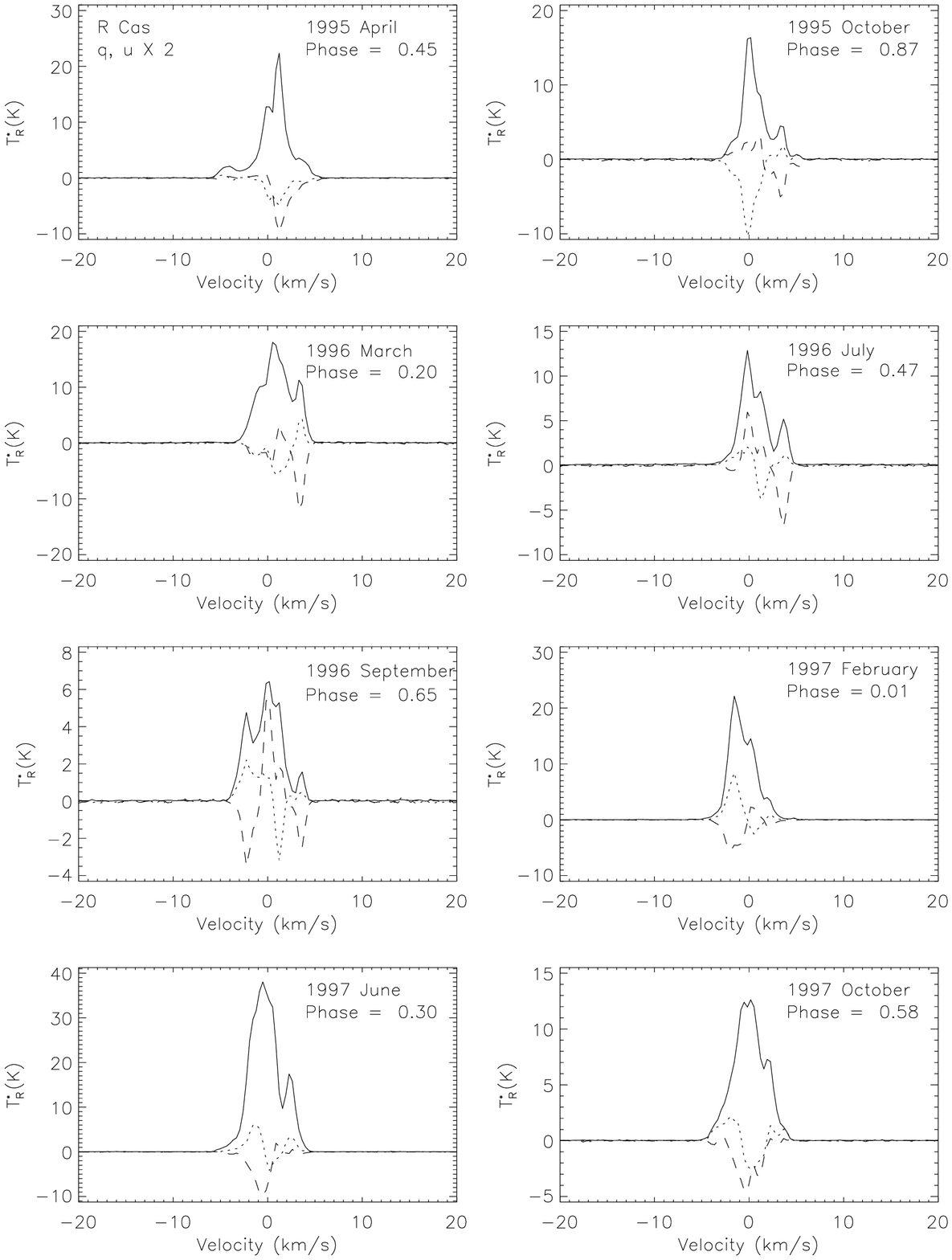}
      \caption{
SiO line temperature and polarization spectra for R Cas.  The
non-normalized Stokes parameters, $q$ (dashed lines) and $u$ (dotted
lines), have been multiplied by two in all of the panels.}
    \end{figure}
    \placefigure{fig:R_Cas}

\clearpage

    \begin{figure}
      \includegraphics[scale=0.8]{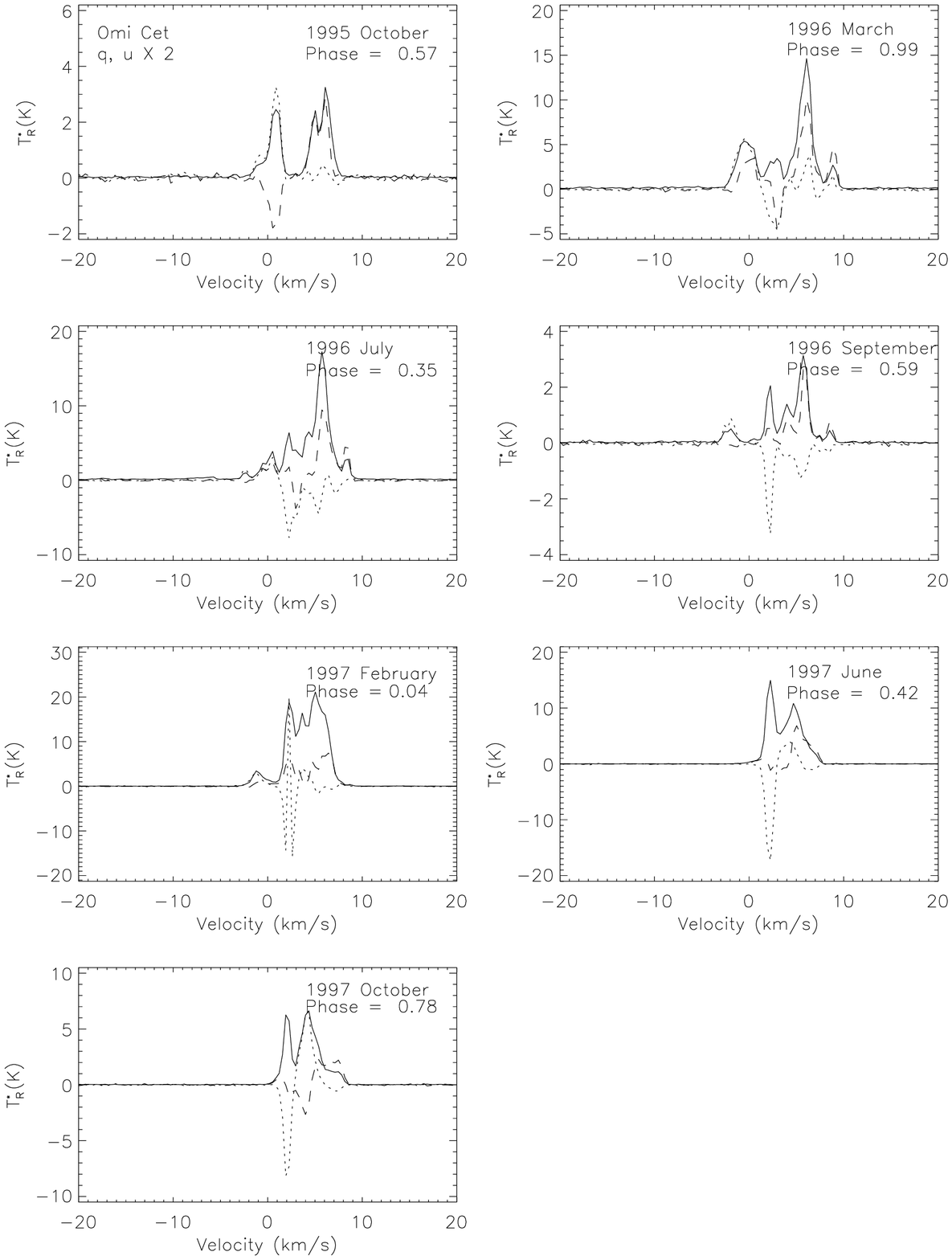}
      \caption{
SiO line temperature and polarization spectra for Omi Ceti.  The
non-normalized Stokes parameters, $q$ (dashed lines) and $u$ (dotted
lines), have been multiplied by two in all of the panels.}
    \end{figure}
    \placefigure{fig:Omi_Cet}

\clearpage

    \begin{figure}
      \includegraphics[scale=0.8]{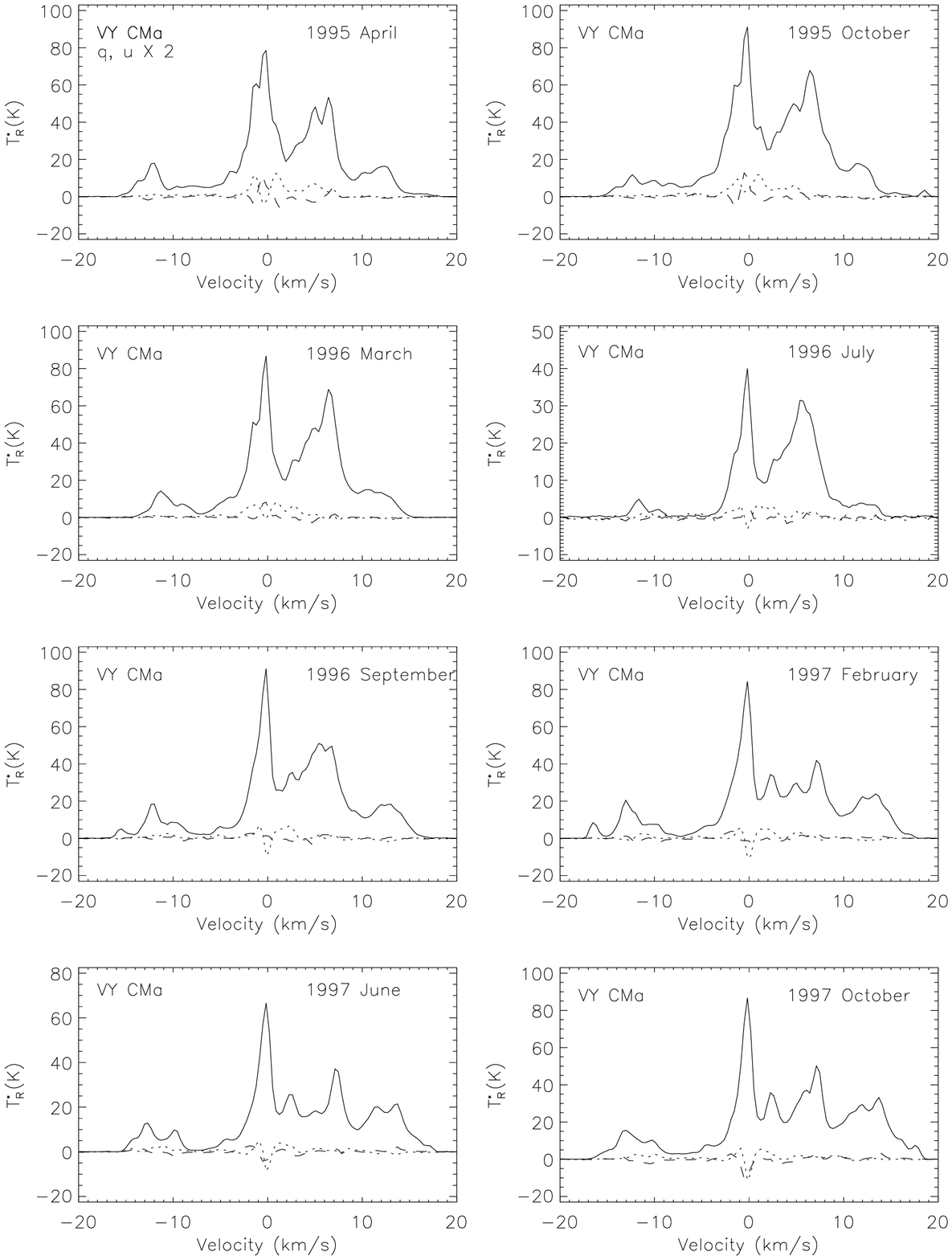}
      \caption{
SiO line temperature and polarization spectra for VY CMa.  The
non-normalized Stokes parameters, $q$ (dashed lines) and $u$ (dotted
lines), have been multiplied by two in all of the panels.}
    \end{figure}
    \placefigure{fig:VY CMa}

\clearpage

    \begin{figure}
      \includegraphics[scale=0.8]{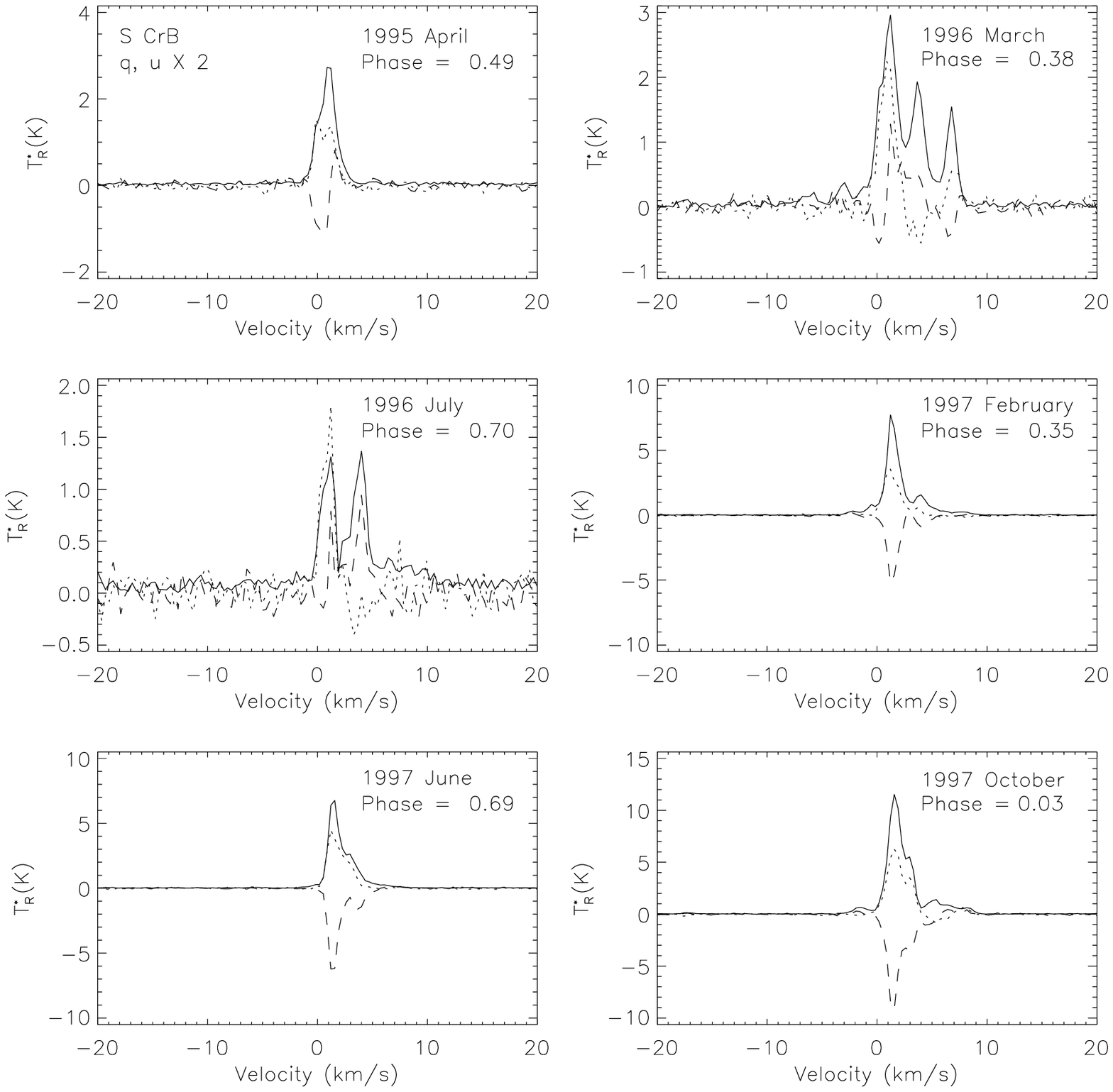}
      \caption{
SiO line temperature and polarization spectra for S CrB.  The
non-normalized Stokes parameters, $q$ (dashed lines) and $u$ (dotted
lines), have been multiplied by two in all of the panels.}
    \end{figure}
    \placefigure{fig:S_CrB}

\clearpage

    \begin{figure}
      \includegraphics[scale=0.8]{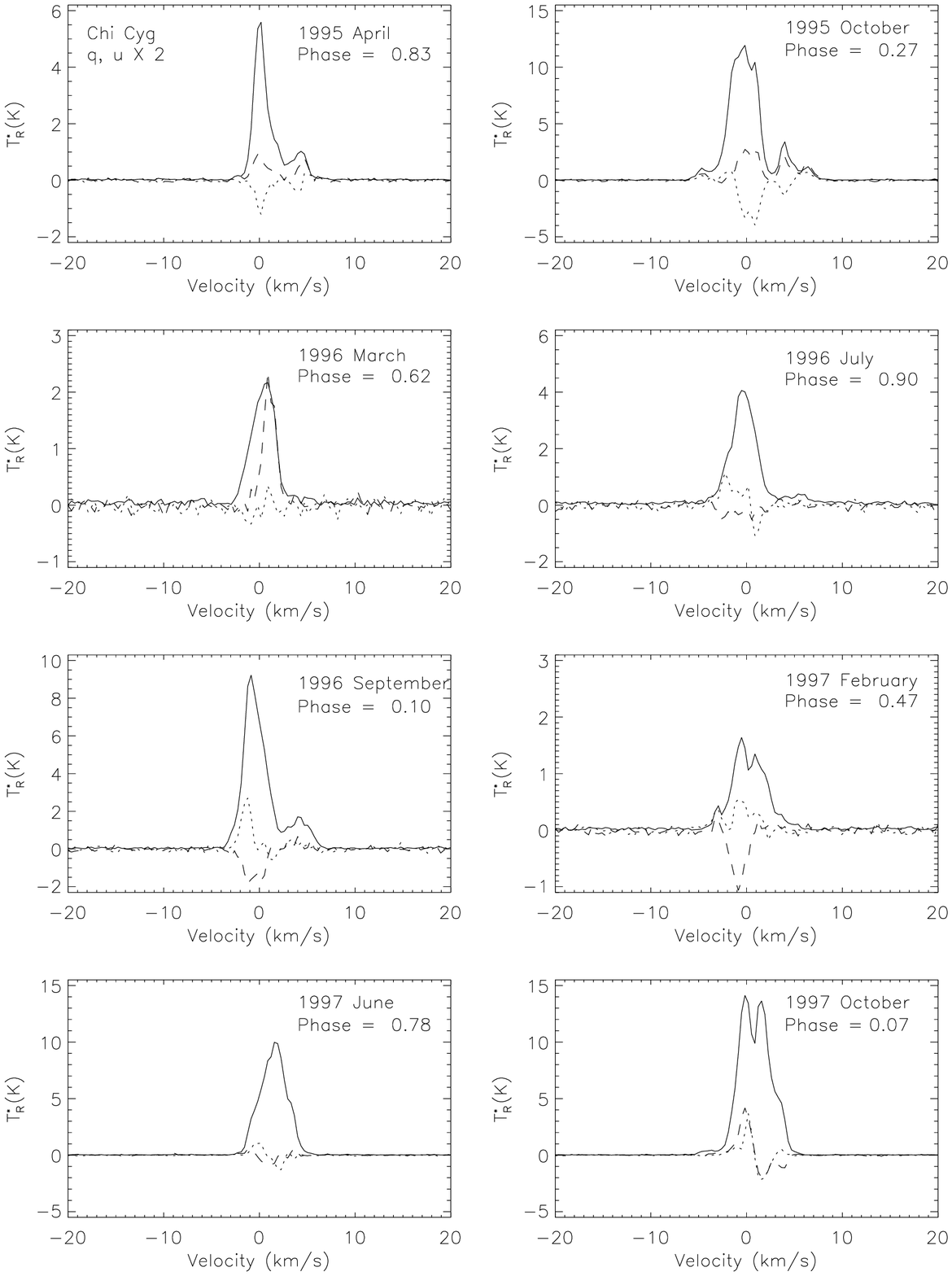}
      \caption{
SiO line temperature and polarization spectra for Chi Cyg.  The
non-normalized Stokes parameters, $q$ (dashed lines) and $u$ (dotted
lines), have been multiplied by two in all of the panels.}
    \end{figure}
    \placefigure{fig:Chi Cyg}

\clearpage

    \begin{figure}
      \includegraphics[scale=0.8]{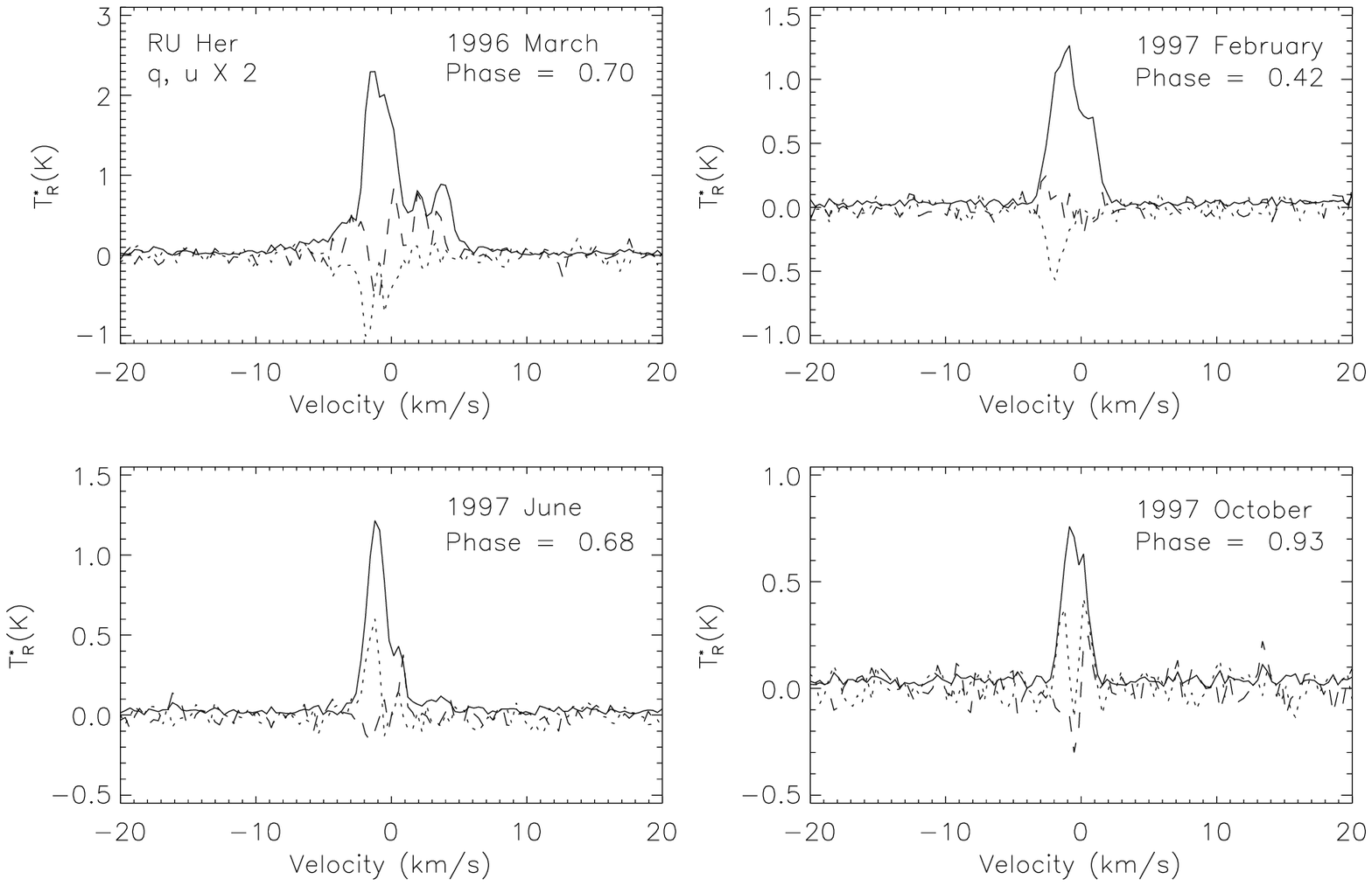}
      \caption{
SiO line temperature and polarization spectra for RU Her.  The
non-normalized Stokes parameters, $q$ (dashed lines) and $u$ (dotted
lines), have been multiplied by two in all of the panels.}
    \end{figure}
    \placefigure{fig:RU_Her}

\clearpage

    \begin{figure}
      \includegraphics[scale=0.8]{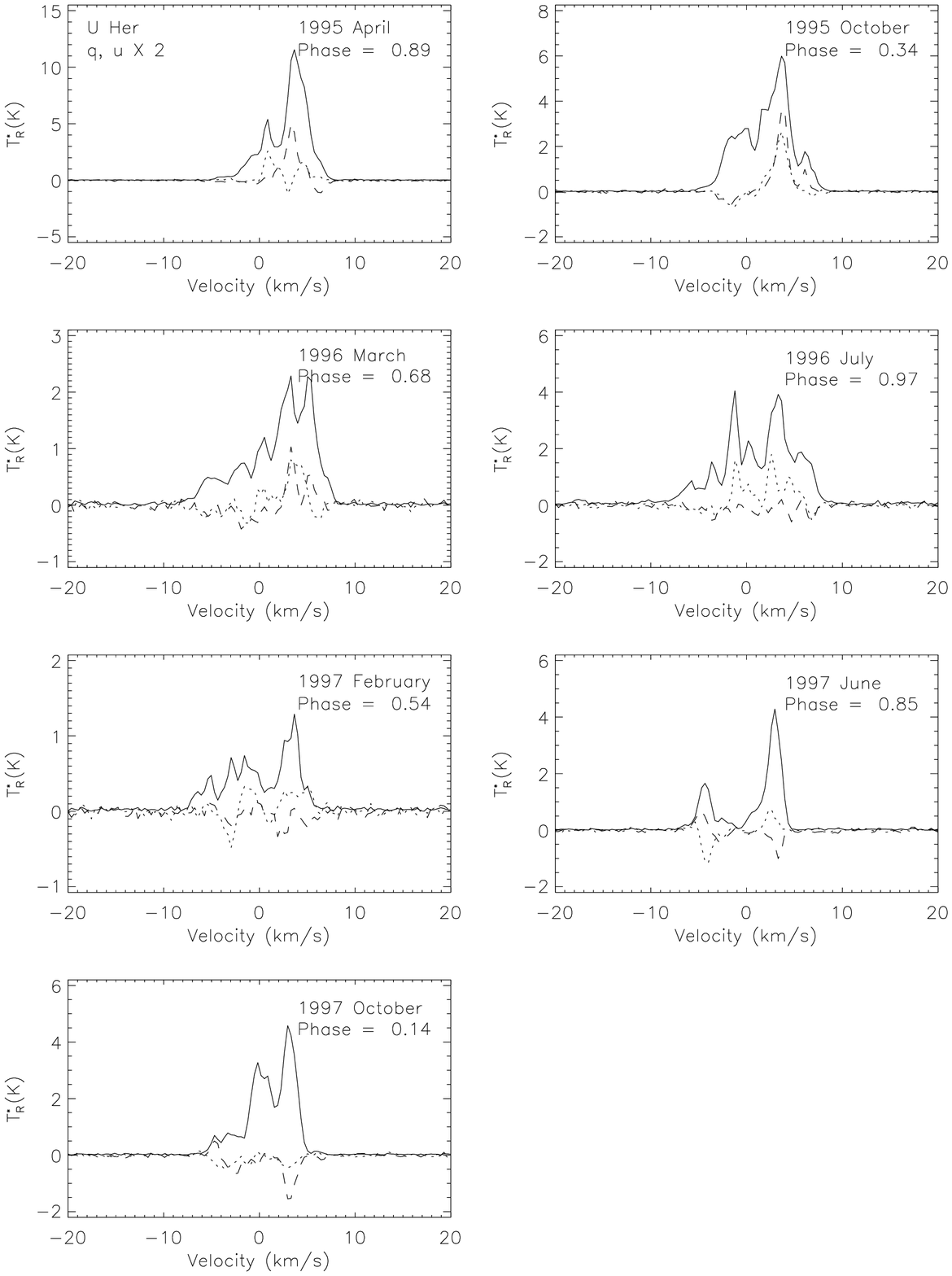}
      \caption{
SiO line temperature and polarization spectra for U Her.  The
non-normalized Stokes parameters, $q$ (dashed lines) and $u$ (dotted
lines), have been multiplied by two in all of the panels.}
    \end{figure}
    \placefigure{fig:U_Her}

\clearpage

    \begin{figure}
      \includegraphics[scale=0.8]{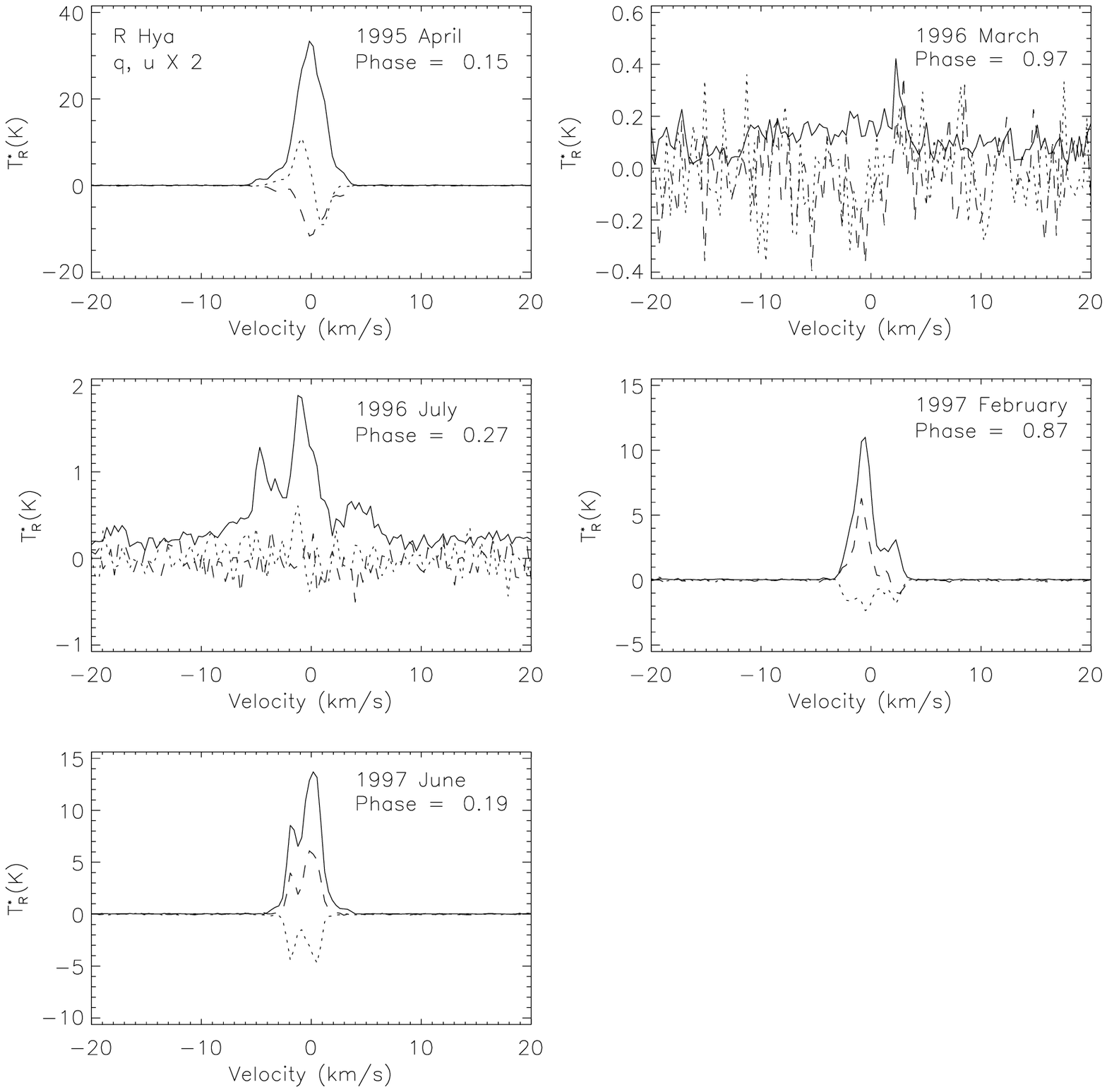}
      \caption{
SiO line temperature and polarization spectra for R Hya.  The
non-normalized Stokes parameters, $q$ (dashed lines) and $u$ (dotted
lines), have been multiplied by two in all of the panels.}
    \end{figure}
    \placefigure{fig:R_Hya}

\clearpage

    \begin{figure}
      \includegraphics[scale=0.8]{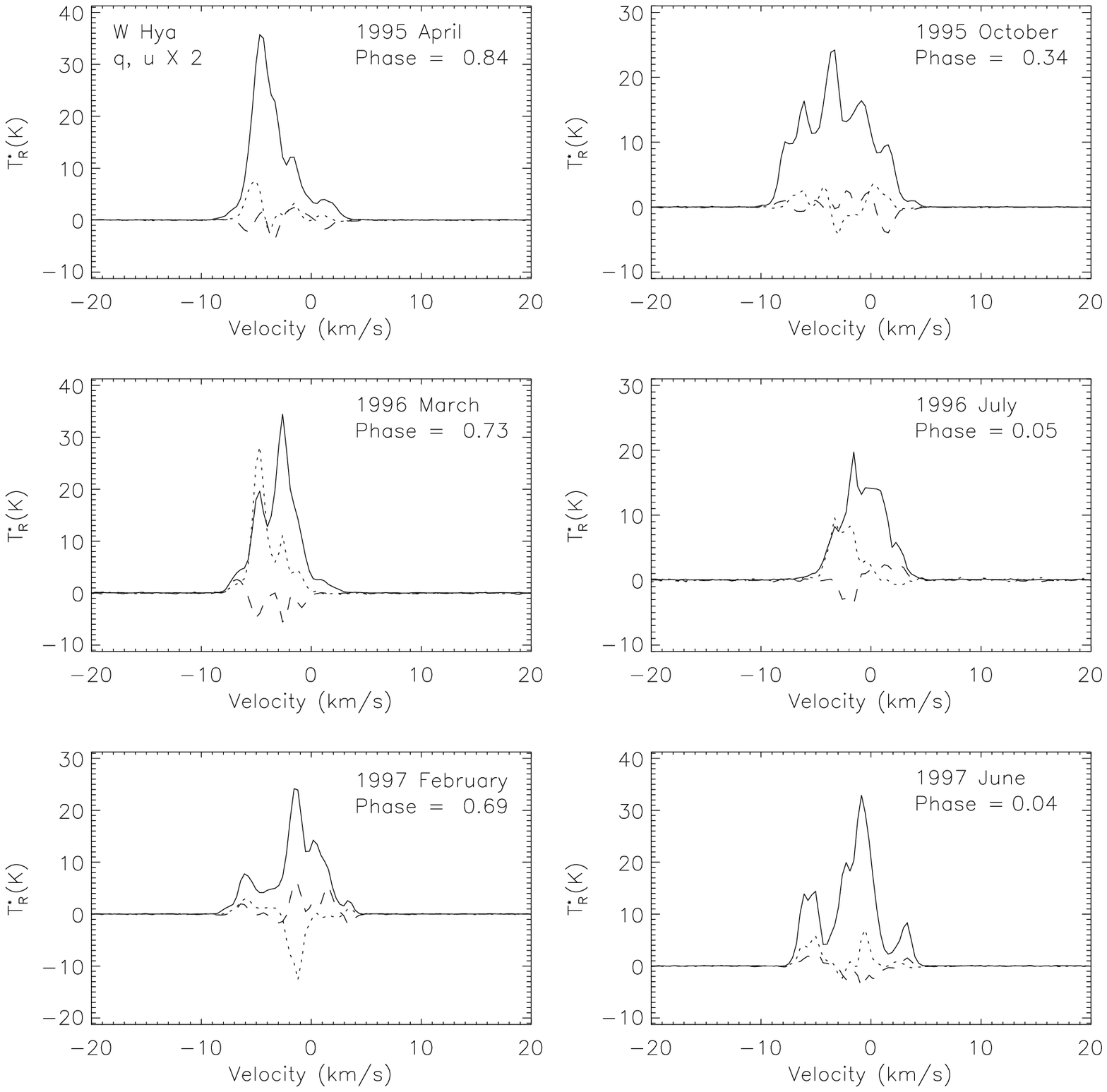}
      \caption{
SiO line temperature and polarization spectra for W Hya.  The
non-normalized Stokes parameters, $q$ (dashed lines) and $u$ (dotted
lines), have been multiplied by two in all of the panels.}
    \end{figure}
    \placefigure{fig:W_Hya}

\clearpage

    \begin{figure}
      \includegraphics[scale=0.8]{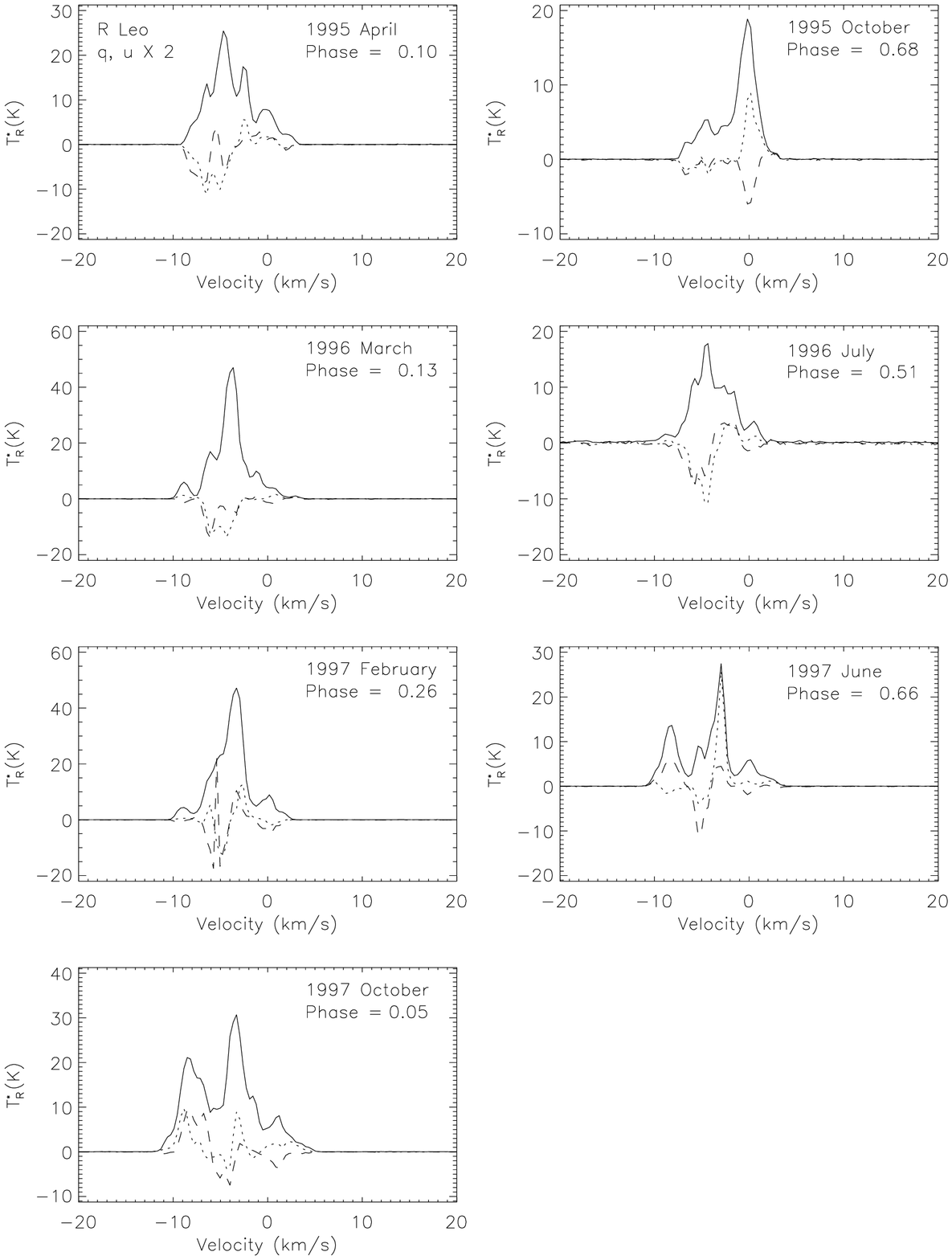}
      \caption{
SiO line temperature and polarization spectra for R Leo.  The
non-normalized Stokes parameters, $q$ (dashed lines) and $u$ (dotted
lines), have been multiplied by two in all of the panels.}
    \end{figure}
    \placefigure{fig:R_Leo}

\clearpage

    \begin{figure}
      \includegraphics[scale=0.8]{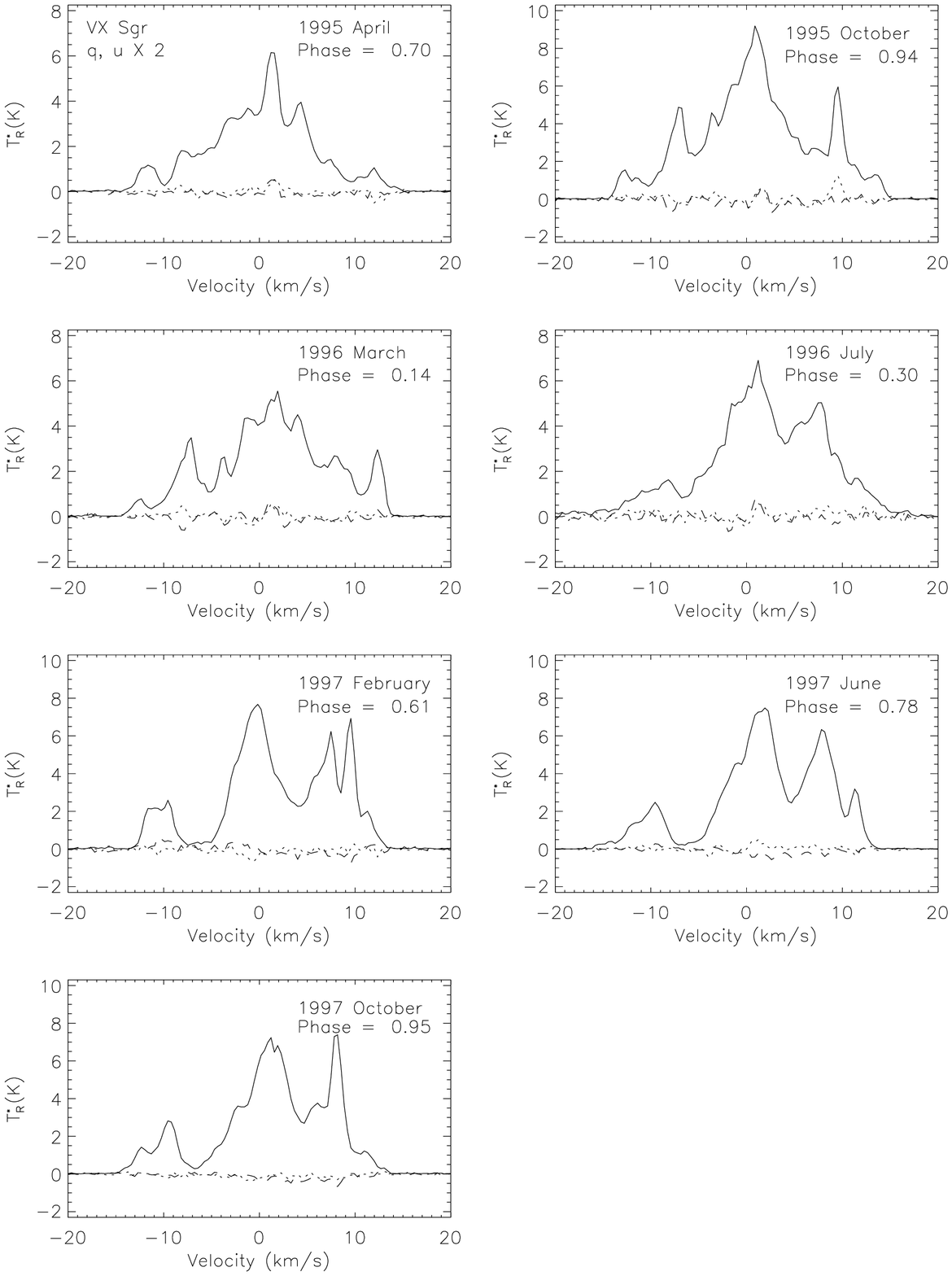}
      \caption{
SiO line temperature and polarization spectra for VX Sgr.  The
non-normalized Stokes parameters, $q$ (dashed lines) and $u$ (dotted
lines), have been multiplied by two in all of the panels.}
    \end{figure}
    \placefigure{fig:VX Sgr}

\clearpage

    \begin{figure}
      \includegraphics[scale=0.8]{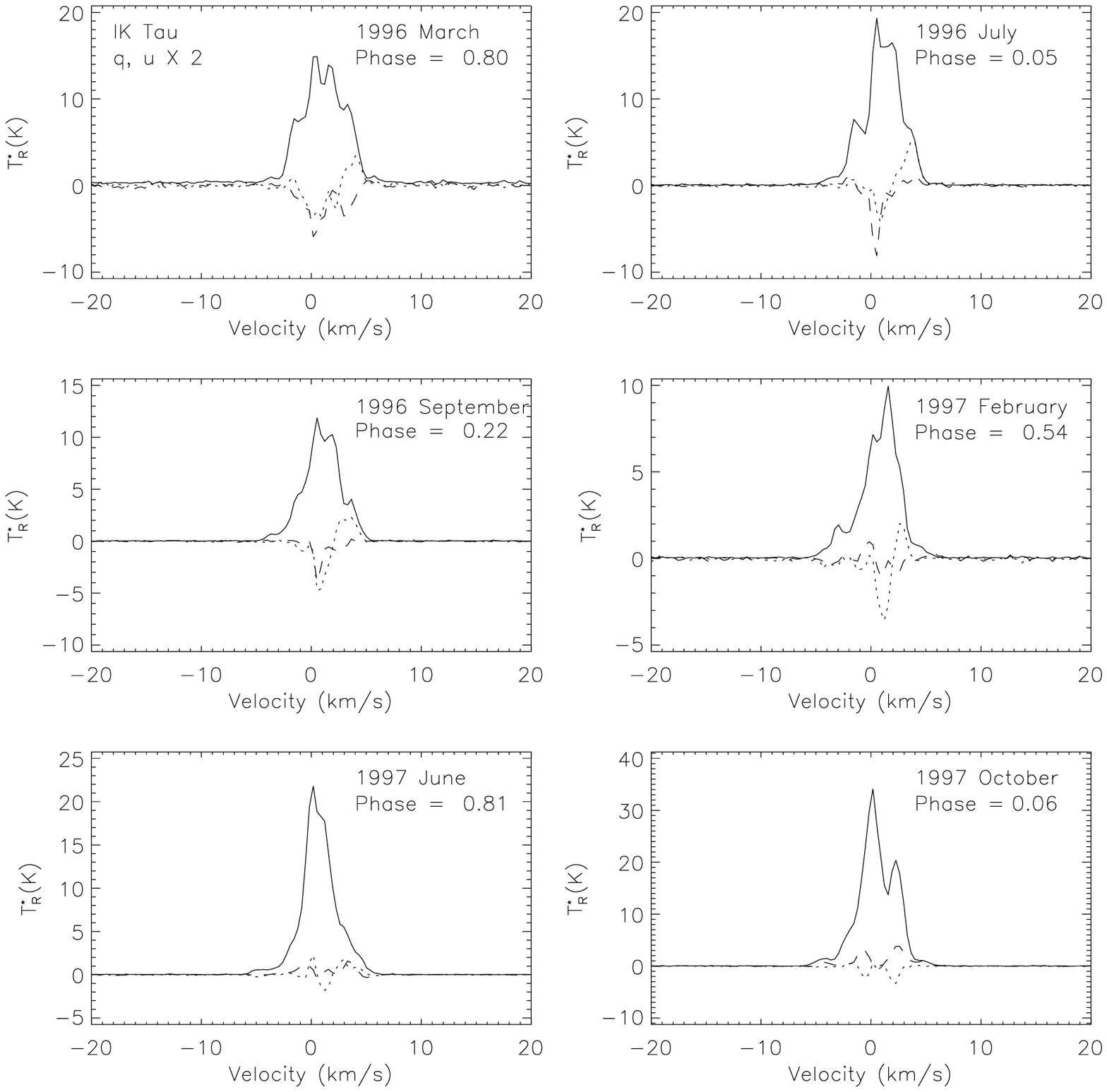}
      \caption{
SiO line temperature and polarization spectra for IK Tau.  The
non-normalized Stokes parameters, $q$ (dashed lines) and $u$ (dotted
lines), have been multiplied by two in all of the panels.}
    \end{figure}
    \placefigure{fig:IK Tau}

\clearpage

    \begin{figure}
      \includegraphics[scale=0.8]{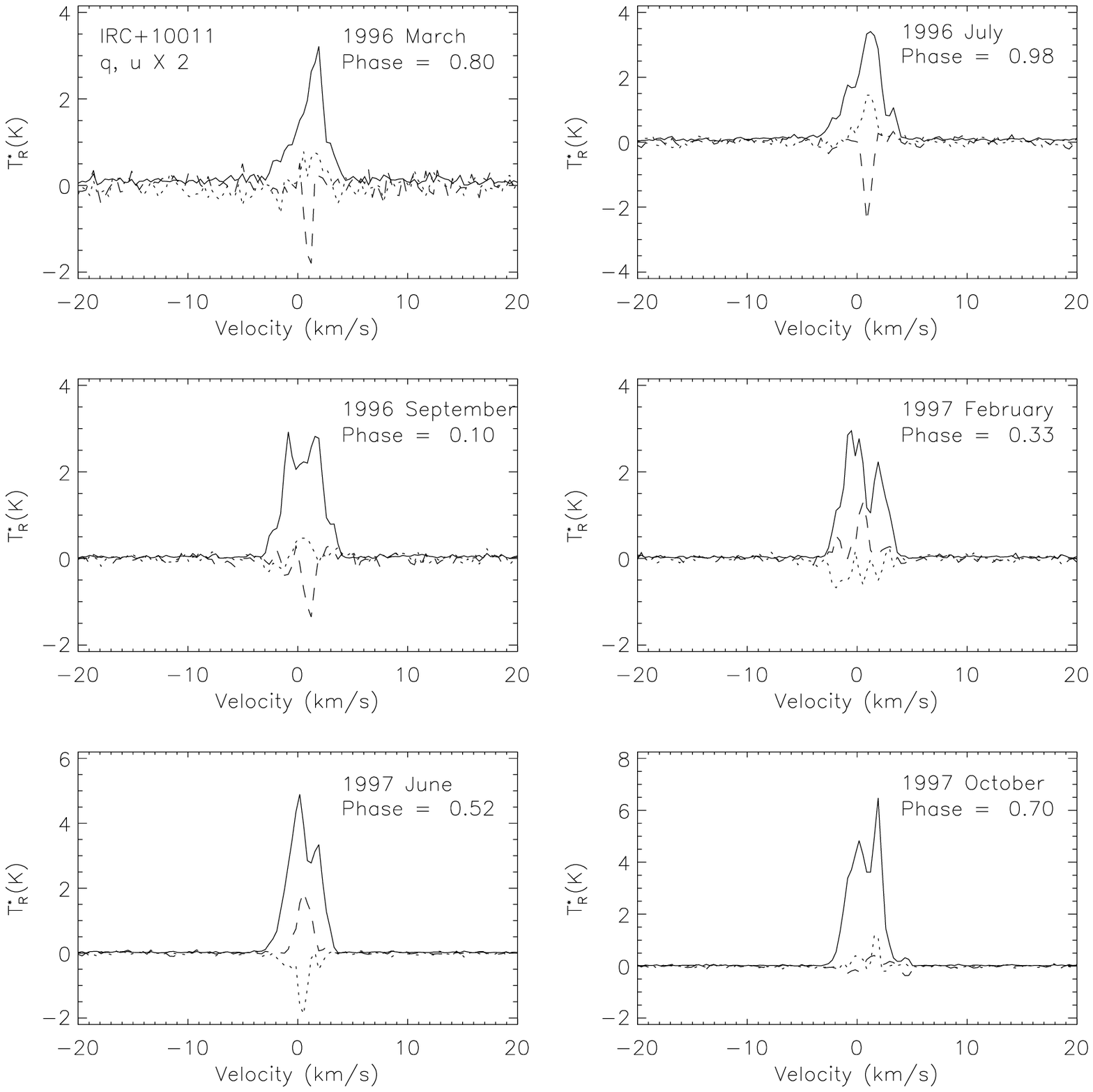}
      \caption{
SiO line temperature and polarization spectra for IRC+10011.  The
non-normalized Stokes parameters, $q$ (dashed lines) and $u$ (dotted
lines), have been multiplied by two in all of the panels.}
    \end{figure}
    \placefigure{fig:IRC10011}

\clearpage

    \begin{figure}
      \includegraphics[scale=0.8]{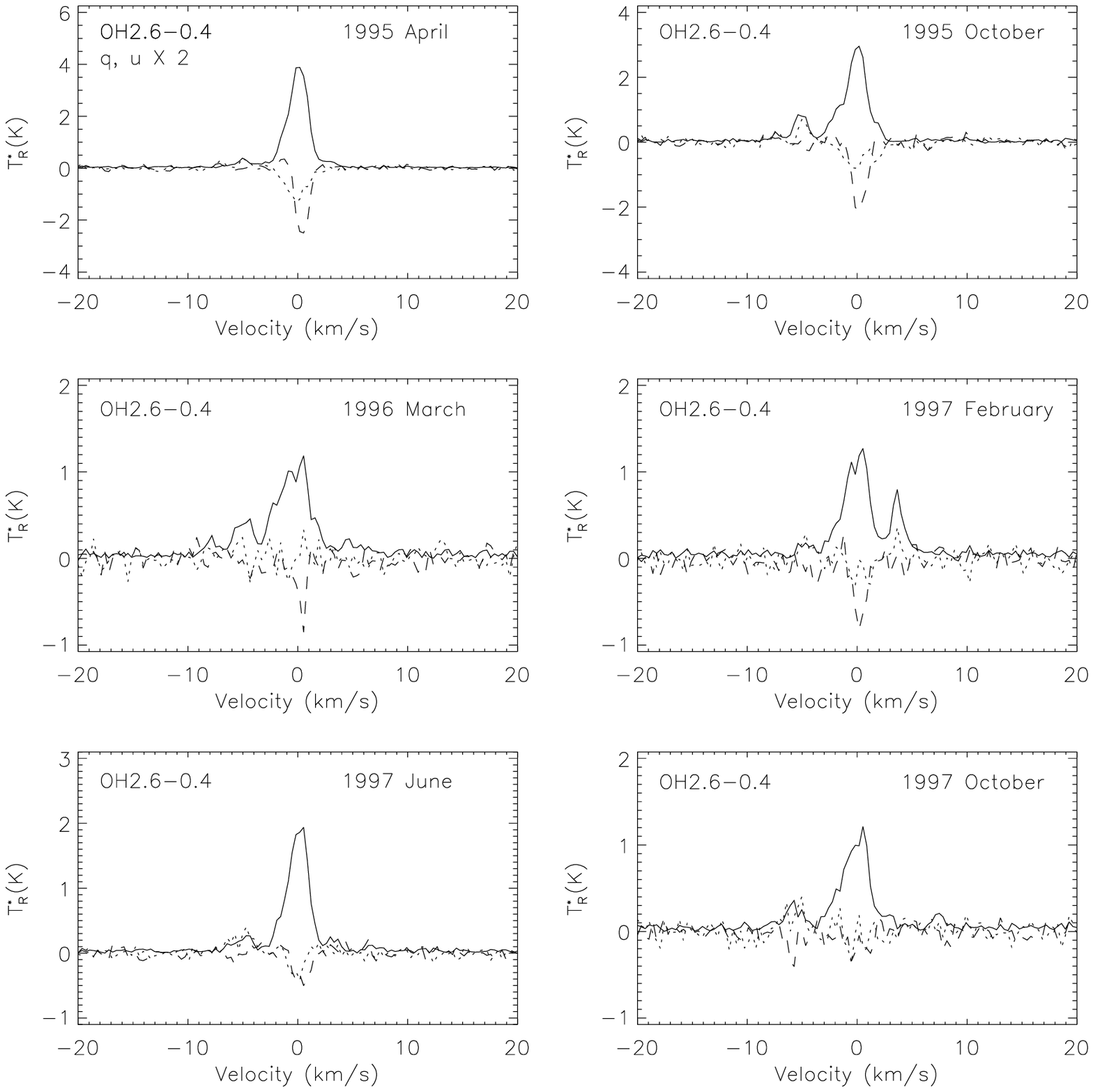}
      \caption{
SiO line temperature and polarization spectra for OH2.6-0.4.  The
non-normalized Stokes parameters, $q$ (dashed lines) and $u$ (dotted
lines), have been multiplied by two in all of the panels.}
    \end{figure}
    \placefigure{fig:OH2.6-0.4}

\end{document}